\documentclass[a4paper,UKenglish,cleveref, autoref, thm-restate]{lipics-v2021}

\usepackage{url}

\usepackage{tikz,cite}
\usetikzlibrary{positioning,fit,calc,shapes,backgrounds}
\usetikzlibrary{matrix}
\usetikzlibrary{arrows}
\usetikzlibrary{arrows.meta}
\usetikzlibrary{decorations.markings}

\usepackage{amsmath,amssymb}
\usepackage{subcaption}
\usepackage{mathtools}

\usepackage{complexity}
\usepackage{environ,multirow}
\usepackage{setspace}
\usepackage[normalem]{ulem}

\usepackage{cancel}

\usepackage{hyperref,xspace}

\newcommand{\Real}{{\mathbb R}}

\makeatletter
\tikzset{
    dot diameter/.store in=\dot@diameter,
    dot diameter=3pt,
    dot spacing/.store in=\dot@spacing,
    dot spacing=10pt,
    dots/.style={
        line width=\dot@diameter,
        line cap=round,
        dash pattern=on 0pt off \dot@spacing
    },
}
\makeatother

\title{Maximum cut on interval graphs of interval count four is $\NP$-complete} 

\author{Celina M. H. de Figueiredo}{Federal University of Rio de Janeiro, Rio de Janeiro, Brazil}{celina@cos.ufrj.br}{https://orcid.org/0000-0002-6393-0876}{}

\author{Alexsander A. de Melo}{Federal University of Rio de Janeiro, Rio de Janeiro, Brazil}{aamelo@cos.ufrj.br}{https://orcid.org/0000-0001-5268-6997}{} 

\author{Fabiano S. Oliveira}{Rio de Janeiro State University, Rio de Janeiro, Brazil}{fabiano.oliveira@ime.uerj.br}{https://orcid.org/0000-0002-8498-2472}{} 

\author{Ana Silva}{Federal University of Cear\'{a}, Cear\'{a}, Brazil}{anasilva@mat.ufc.br}{https://orcid.org/0000-0001-8917-0564}{} 

\authorrunning{C.\,M.\,H. de Figueiredo et al.} 

\Copyright{Celina M. H. de Figueiredo, Alexsander A. de Melo, Fabiano S. Oliveira, and Ana Silva}



\category{} 

\relatedversion{}

\acknowledgements{This project was partially supported by CNPq, FAPERJ and CNPq/FUNCAP.}

\nolinenumbers

\EventEditors{}
\EventNoEds{2}
\EventLongTitle{}
\EventShortTitle{}
\EventAcronym{}
\EventYear{}
\EventDate{}
\EventLocation{}
\EventLogo{}
\SeriesVolume{}
\ArticleNo{}

\begin{document}

\maketitle

\abstract{The computational complexity of the \textsc{MaxCut} problem restricted to interval graphs has been open since the 80's, being one of the problems proposed by Johnson in his \emph{Ongoing Guide to NP-completeness}, and has been settled as NP-complete only recently by  Adhikary, Bose, Mukherjee and Roy. On the other hand, many flawed proofs of polynomiality for \textsc{MaxCut} on the more restrictive class of unit/proper interval graphs (or graphs with interval count~1) have been presented along the years, and the classification of the problem is still unknown. In this paper, we present the first NP-completeness proof for \textsc{MaxCut} when restricted to interval graphs with bounded interval count, namely graphs with interval count~4. 
}

\section{Introduction}
\label{s:intro}
A \emph{cut} is a partition of the vertex set of a graph into two disjoint parts and the \emph{maximum cut problem} (denoted \textsc{MaxCut} for short) aims to determine a cut with the maximum number of edges for which each endpoint is in a distinct part. 
The decision problem \textsc{MaxCut} is known to be NP-complete since the seventies~\cite{GJS76}, 
and only recently its restriction to interval graphs has been announced to be hard~\cite{ABMR20}, settling a long-standing open problem that appeared in a summary table in the 1985 column of the  {\it Ongoing Guide to NP-completeness} by David S. Johnson~\cite{J85}.
We refer the reader to a revised version of Johnson's summary  table in~\cite{FMSSarxiv}, where one can also find a parameterized complexity version of the said table.

An \emph{interval model} is a family of closed intervals of the real line. 
A graph is an \emph{interval graph} if there exists an interval model, for which each interval corresponds to a vertex of the graph, such that distinct vertices are adjacent in the graph if and only if the corresponding intervals intersect. 
Ronald L. Graham proposed in the 80's the study of the \emph{interval count} of an interval graph as the smallest number of interval lengths used by an interval model of the graph. Interval graphs having interval count $1$ are called \emph{unit interval graphs} (these are also called proper interval graphs, or indifference graphs). 
Understanding 
the interval count,
besides being an interesting and challenging problem by itself, can be also of value for the investigation of problems that are hard for general interval graphs, and easy for unit interval graphs (e.g. geodetic number~\cite{EEH.12,CDFGLR.20}, optimal linear arrangement~\cite{CFHKK.06,JZ.95}, sum coloring~\cite{NSS.99,M.05}). The positive results for unit interval graphs usually take advantage of the fact that a representation for these graphs can be found in linear time~\cite{CKNOS.95,FMM.95}. 
Surprisingly, the recognition of interval graphs with interval count $k$ is open, even for $k=2$~\cite{COS2012}. Nevertheless, another generalization of unit interval graphs has been recently introduced which might be more promising in this aspect. 
These graphs are 
called \emph{$k$-nested interval graphs}, 
for which an efficient recognition algorithm has firstly appeared in~\cite{COS2011}. Recently, a linear time algorithm has been devised in~\cite{KOS2018}.

In the same way that \textsc{MaxCut} on interval graphs has evaded being solved for so long, the community has been puzzled by the restriction to unit interval graphs. Indeed, two attempts at solving it in polynomial time were proposed in~\cite{BKN99,BES17} just to be 
disproved closely after~\cite{WEA04,KMN20}. In this paper, we give the first classification that bounds the interval count, namely, we prove that \textsc{MaxCut} is NP-complete when restricted to interval graphs of interval count~4. 
This also implies NP-completeness 
for the newly generalized class of $4$-nested graphs, 
and opens the search for a full polynomial/NP-complete dichotomy classification in terms of the interval count. It can still happen that the problem is hard even on graphs of interval count~1.
We contribute towards filling the complexity gap between interval and unit interval graphs.
We have communicated the result at the MFCS 2021 conference~\cite{FMOS21}, and previous versions of the full proof appeared in the ArXiv~\cite{FMOS20}. The present paper contains the improved and much shorter full proof.

Next, we establish basic definitions and notation. 
Section~\ref{s:redu} describes our reduction and Section~\ref{s:orig} discusses  the interval count of the interval graph constructed in~\cite{ABMR20}.

\subsection{Preliminaries}
\label{s:prim}
In this work, all graphs considered are simple. 
For missing definitions and notation of graph theory, we refer to~\cite{BM2008}. 
For a comprehensive study of interval graphs, we refer to~\cite{FIS85}.

Let $G$ be a graph. 
Let $X$ and $Y$ be two disjoint subsets of $V(G)$. 
We let $E_G(X,Y)$ be the set of edges of $G$ with an endpoint in $X$ and the other endpoint in $Y$. 
A \emph{cut} of $G$ is a partition of $V(G)$ into two parts $A, B \subseteq V(G)$, denoted by $[A, B]$; 
the edge set $E_G(A,B)$ is called the \emph{cut-set} of $G$ associated with $[A,B]$. 
The \emph{size of a cut-set} is defined as its cardinality.
The \emph{size of a cut} is the size of its associated cut-set.
For each two vertices $u,v \in V(G)$, we say that $u$ and $v$ \emph{are in a same part of $[A,B]$} if either $\{u,v\} \subseteq A$ or $\{u,v\} \subseteq B$; otherwise, we say that $u$ and $v$ \emph{are in opposite parts of $[A,B]$}. 
Denote by $\mathsf{mc}(G)$ the maximum size of a cut-set of $G$. 
The \textsc{MaxCut} problem has as input a graph $G$ and a positive integer $k$, and it asks whether $\mathsf{mc}(G) \geq k$.

Let $I \subseteq \Real$ be a closed interval of the real line.
We let $\ell(I)$ and $r(I)$ denote respectively the minimum and maximum points of $I$, which we call the \emph{left} and the \emph{right endpoints} of $I$, respectively. 
For every non-empty collection of intervals $\cal{H}$, we define the \emph{left endpoint} of $\mathcal{H}$ as $\ell(\mathcal{H}) = \min_{I \in \mathcal{H}} \ell(I)$ and the \emph{right endpoint} of $\mathcal{H}$ as $r(\mathcal{H}) = \max_{I \in \mathcal{H}} r(I)$. 
We denote a closed interval $I$ by $[\ell(I),r(I)]$. Distinction from the cut notation will be clear from the context. For every two intersecting intervals $I$ and $I'$, we say that $I$ \emph{covers} $I'$ if $\ell(I) \le  \ell(I')$ and $r(I) \ge r(I')$,  that $I$ \emph{intersects $I'$ to the left} if $\ell(I) < \ell(I') < r(I) < r(I')$, and that $I$ \emph{intersects $I'$ to the right} if $\ell(I') < \ell(I) < r(I) < r(I)$. 
We say that an interval $I$ \emph{precedes} an interval $I'$ if $r(I) < \ell(I')$; and more generally, we say that a collection of intervals $\mathcal{H}$ \emph{occurs to the left} of a collection $\mathcal{H}'$ if every interval in $\mathcal{H}$ precedes every interval in $\mathcal{H}'$.  
The \emph{length} of an interval $I$ is defined as $\lvert I\rvert  =r(I) - \ell(I)$. 

An \emph{interval model} is a finite multiset ${\cal M}$ of intervals. 
The \emph{interval count} of an interval model ${\cal M}$, denoted by $\mathsf{ic}({\cal M})$, is defined as the number of distinct lengths of the intervals in ${\cal M}$. 
Let $G$ be a graph and ${\cal M}$ be an interval model. 
An \emph{${\cal M}$-representation} of $G$ is a bijection $\phi \colon V(G) \rightarrow {\cal M}$ such that, for every two distinct vertices $u, v \in V(G)$, we have that $uv \in E(G)$ if and only if $\phi(u) \cap \phi(v) \neq \emptyset$. 
If such an ${\cal M}$-representation exists, we say that ${\cal M}$ is an \emph{interval model of $G$}. 
We note that a graph may have either no interval model or arbitrarily many distinct interval models. 
A graph is called an \emph{interval graph} if it has an interval model. 
The \emph{interval count of an interval graph~$G$}, denoted by $\mathsf{ic}(G)$, is defined as 
$\mathsf{ic}(G) = \min\{\mathsf{ic}({\cal M}) \colon {\cal M} \text{ is an interval model of } G\}\text{.}$  
An interval graph is called a \emph{unit interval graph} if its interval count is equal to $1$.

Note that, for every interval model ${\cal M}$, there exists a unique (up to isomorphism) graph that admits an ${\cal M}$-representation. 
Thus, for every interval model ${\cal M}=\{I_{1},\ldots,I_{n}\}$, we let $\mathbb{G}_{{\cal M}}$ be the graph with vertex set $V(\mathbb{G}_{{\cal M}}) = \{1,\ldots,n\}$ and edge set $E(\mathbb{G}_{{\cal M}}) = \{ij \colon I_{i}, I_{j} \in {\cal M},\, I_{i} \cap I_{j} \neq \emptyset,\, i\neq j\}$. 
Since $\mathbb{G}_{{\cal M}}$ is uniquely determined (up to isomorphism) from ${\cal M}$, in what follows we may make an abuse of language and use graph terminologies to describe properties related to the intervals in ${\cal M}$. 
Two intervals $I_i, I_j \in {\cal M}$ are said to be \emph{true twins} in $\mathbb{G}_{{\cal M}}$ if they have the same closed neighborhood in $\mathbb{G}_{{\cal M}}$. 

\medskip

\section{Our reduction}\label{s:redu}

The following theorem is the main  contribution of this work: 

\begin{theorem}\label{theo:main_result}
    \textsc{MaxCut} is NP-complete on interval graphs of interval count~$4$.
\end{theorem}
This result is a stronger version of that of Adhikary et al.~\cite{ABMR20}. To prove Theorem~\ref{theo:main_result}, we present a polynomial-time reduction from \textsc{MaxCut} on cubic graphs, which is known to be NP-complete~\cite{BK99}. Since our proof is based on that of Adhikary et al., we start by presenting some important properties of their key gadget.

\smallskip

 \subsection{Grained gadget}
 \label{ss:grai}

The interval graph constructed in the reduction of~\cite{ABMR20} is strongly based on two types of gadgets, called \emph{V-gadgets} and \emph{E-gadgets}. 
In fact, these gadgets have the same structure except for the number of intervals of certain kinds contained in each of them. 
In this subsection, we present a generalization of such gadgets, rewriting their key properties to suit our purposes. 
In order to discuss the interval count of the reduction of~\cite{ABMR20}, we describe it in detail in Section~\ref{s:orig}.

\begin{figure}[ht]\centering
	\includegraphics[scale=4.5]{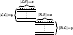}
	\caption{General structure of an $(x,y)$-grained gadget.}\label{fig:grainedgadget}
\end{figure}

Let $x$ and $y$ be two positive integers. 
An \emph{$(x,y)$-grained gadget} (see Figure~\ref{fig:grainedgadget} to follow) is an interval model $\cal H$ formed by $2y$ long intervals, $y$ of which called \emph{left long} and $y$ called \emph{right long} intervals, together with $2x$ pairwise disjoint short intervals, $x$ of which called \emph{left short} 
and $x$ of which called \emph{right short}. 
The $y$ left long intervals all have the same right endpoint, which also is the left endpoint of each of the $y$ right long intervals. The $x$ left (resp. right) short intervals are all pairwise disjoint and intersect each left (resp. right) long interval, but intersect no right (resp. left) long interval. 
We write ${\cal LS}({\cal H})$, ${\cal LL}({\cal H})$, ${\cal RS}({\cal H})$ and ${\cal RL}({\cal H})$ to denote the left short, left long, right short and right long intervals of ${\cal H}$, respectively. And we omit ${\cal H}$ when it is clear from the context.

Note that, if ${\cal H}$ is an $(x,y)$-grained gadget, then $\mathbb{G}_{{\cal H}}$ is a split graph such that ${\cal LS} \cup {\cal RS}$ is an independent set of size $2x$, ${\cal LL} \cup {\cal RL}$ is a clique of size~$2y$, 
and, for every vertex $u \in {\cal LS}$, $N_{\mathbb{G}_{{\cal H}}}(u) = {\cal LL}$
and, for every vertex $u \in {\cal RS}$, $N_{\mathbb{G}_{{\cal H}}}(u) = {\cal RL}$.
Moreover, the intervals in ${\cal LL}$ are true twins in $\mathbb{G}_{{\cal H}}$; similarly, the intervals in ${\cal RL}$ are true twins in $\mathbb{G}_{{\cal H}}$. 

\begin{figure}[ht]\centering\captionsetup[subfigure]{justification=centering}
	\begin{subfigure}[t]{0.32\textwidth}\centering
		\includegraphics[scale = 2.8]{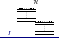}
		\caption{Cover intersection}\label{subfig:fig1_0}
	\end{subfigure}
	\begin{subfigure}[t]{0.32\textwidth}\centering
		\includegraphics[scale = 2.8]{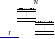}
		\caption{Weak intersection to the left}\label{subfig:fig1_1}
	\end{subfigure}
\begin{subfigure}[t]{0.32\textwidth}\centering
		\includegraphics[scale = 2.8]{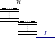}
		\caption{Weak intersection to the right}\label{subfig:fig1_2}
	\end{subfigure}
	\vspace{2.5ex}
	
\begin{subfigure}[t]{0.44\textwidth}\centering
		\includegraphics[scale = 2.8]{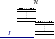}
		\caption{Strong intersection to the left}\label{subfig:fig1_3}
	\end{subfigure}
\begin{subfigure}[t]{0.44\textwidth}\centering
		\includegraphics[scale = 2.8]{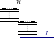}
		\caption{Strong intersection to the right}\label{subfig:fig1_4}
	\end{subfigure}
	\caption{Interval $I \in {\cal M} \setminus {\cal H}$
	(\protect\subref{subfig:fig1_0}) covering ${\cal H}$, (\protect\subref{subfig:fig1_1}-\protect\subref{subfig:fig1_2}) weakly intersecting ${\cal H}$ to the left and to the right, and (\protect\subref{subfig:fig1_3}-\protect\subref{subfig:fig1_4}) strongly intersecting ${\cal H}$ to the left and to the right.}
	\label{fig:fig1}
\end{figure}

Let ${\cal M}$ be an interval model containing an $(x,y)$-grained gadget ${\cal H}$. 
We say that an interval of ${\cal M}\setminus  {\cal H}$ \emph{intersects} ${\cal H}$ if it intersects at least one interval of ${\cal H}$. Otherwise, we say that the interval does not intersect ${\cal H}$.
The possible types of intersections between an interval $I\in {\cal M}\setminus  {\cal H}$ and ${\cal H}$ in our construction are depicted in~Figure~\ref{fig:fig1}, 
with the used nomenclature. 
More specifically, the intersection between $I$ and ${\cal H}$ is a \emph{cover intersection} if $I$ intersects all the intervals of ${\cal H}$ (Figure~\ref{fig:fig1}(a)), a \emph{weak intersection to the left} (\emph{right}) if $I$ intersects exactly the left (right) long intervals of ${\cal H}$ (Figure~\ref{fig:fig1}(b–c)), and a \emph{strong intersection to the left} (\emph{right}) if $I$ intersects exactly the left (right) long and short intervals of ${\cal H}$ (Figure~\ref{fig:fig1}(d–e)). 
We say that ${\cal M}$ \emph{respects the structure} of ${\cal H}$ 
if, for every interval $I \in {\cal M}\setminus {\cal H}$, we have that $I$ either does not intersect ${\cal H}$, or
the intersection between $I$ and ${\cal H}$ is of one of the types described above.

The advantage of this gadget is that, by manipulating the values of $x$ and $y$, we can ensure that, in a maximum cut, the left long and right short intervals are placed in the same part, opposite to the part containing the left short and right long intervals, as proved in Lemma~\ref{lemma:short_and_long_partition}, presented shortly.  Note that if ${\cal M}$ is an interval model containing a grained gadget ${\cal H}$ and ${\cal M}$ respects the structure of ${\cal H}$, then every left (resp. right) short interval of ${\cal H}$ intersects exactly the same set of intervals in ${\cal M}$. The following remark will be useful throughout the text.

\begin{remark}
\label{r:half}
Let $[A, B]$ be a maximum cut of a graph $G$.
For any vertex $u \in  V (G)$, if more than half of the neighbours of $u$ are in one part of $[A, B]$, say $A$, then $u \not \in A$, or in other words $u \in B$.
\end{remark}
\begin{proof}
   Suppose that $u \in A$, and let $[A', B']$ be the cut of $G$ such that $A' = A\setminus \{u\}$ and $B' = B \cup \{u\}$. 
   Note that, if $e \in E_{G}(A, B)\setminus E_{G}(A', B')$, then $e$ is incident to~$u$. Thus, since $u$ has more than half of its neighbours in $A$, the size of $[A', B']$ is strictly greater than the size of $[A,B]$, contradicting the maximality of $[A,B]$.
\end{proof}

\begin{lemma}\label{lemma:short_and_long_partition}
Let $x$ and $y$ be positive integers and ${\cal M}$ an interval model containing an $(x,y)$-grained gadget ${\cal H}$. Suppose that ${\cal M}$ respects the structure of ${\cal H}$.
Let $[A, B]$ be a maximum cut of $\mathbb{G}_{\cal M}$.
	Also, let $t$ be the number of intervals in ${\cal M}\setminus{\cal H}$ intersecting ${\cal H}$, 
$\ell$ be the number of intervals in ${\cal M}\setminus{\cal LS}$ intersecting the left short intervals of ${\cal H}$, 
	and $r$ be the number of intervals in ${\cal M\setminus{\cal RS}}$ intersecting the right short intervals of ${\cal H}$. 
	If $\ell$ and $r$ are odd, 
$y > t (x/y -1)$
and $x > t + 2y$,
	then the following hold:
 	\begin{enumerate}
 		\item ${\cal LS}({\cal H}) \subseteq A$ and ${\cal LL}({\cal H}) \subseteq B$, or vice versa;\label{lemma:left_intervals} 
 		\item ${\cal RS}({\cal H}) \subseteq A$ and ${\cal RL}({\cal H}) \subseteq B$, or vice versa; and \label{lemma:right_intervals} 
 		\item  ${\cal LL}({\cal H}) \subseteq A$ and ${\cal RL}({\cal H}) \subseteq B$, or vice versa.\label{lemma:left_long_and_right_long_intervals}
 	\end{enumerate}  
\end{lemma}
\begin{proof}
First, we prove that all the left short intervals are in the same part 
of $[A,B]$. 
Denote by ${\cal N}$ the set of intervals in ${\cal M}\setminus{\cal LS}$ that intersect the left short intervals.

Suppose, without loss of generality, that $B$
contains more than half of the intervals in ${\cal N}$ (it must occur for either $A$ or $B$ since $\ell$ is odd).
Consider any $u \in {\cal LS}$. Then ${\cal N}$ is the set of neighbours of $u$, and since more than half of the intervals of ${\cal N}$ are in $B$, it follows that $u \in A$. This shows that ${\cal LS} \subseteq A$. Thus all the left short intervals are in the same part of $[A, B]$. Because $r$ is also odd, a similar argument shows that all the right short intervals are in the same part of $[A, B]$.

Now consider the left long intervals and suppose, without loss of generality, that all the left short intervals are contained in $A$.
Observe that the number of intervals in ${\cal M}\setminus {\cal LS}$ intersecting a left long interval is 
less than $t+2y < x$. 
Thus every left long interval has more than half of its neighbours from ${\cal LS}$, which are all in one part of $[A,B]$. It now follows that every left long interval is in the part of $[A,B]$ opposite to that of the left short intervals, namely $B$.
An analogous argument holds for the right long intervals. 
This proves Claims (\ref{lemma:left_intervals}) and (\ref{lemma:right_intervals}) in the statement of the lemma.

Finally, let ${\cal L}$ denote the set of long intervals of ${\cal H}$ and suppose by contradiction that ${\cal L}\subseteq A$. 
Let $T$ be the set of intervals in ${\cal M}\setminus {\cal H}$ that intersect ${\cal H}$; then $t=\lvert T \rvert$. Let $t_A =\lvert T \cap A \rvert$ and $t_B =\lvert T \cap B\lvert$. 
Now by switching the intervals in ${\cal RL}$ to $B$ and ${\cal RS}$ to $A$, we gain at least $y^2 + yt_A + xt_B$ cut-edges and lose
at most $xt_A +yt_B$ cut-edges. Since $y > t(x/y - 1)$, we have $y^2 > t(x-y)= xt-yt$ or in other
words, $y^2 > xt_A + xt_B - yt_A - yt_B$. So we get $y^2 + yt_A + xt_B > xt_A + yt_B + 2t_B(x - y)$. As $x > y$, we can conclude that  $y^2 + yt_A + xt_B > xt_A +y t_B$, which means that we have more cut-edges in the new cut than in the cut $[A,B]$, a contradiction.
\end{proof}

We say that $({\cal H},{\cal M})$ is \emph{well-valued} if the conditions of Lemma~\ref{lemma:short_and_long_partition} are satisfied.
Moreover, we say that the constructed model ${\cal M}$ is \emph{well-valued} if all its grained gadgets ${\cal H}$ are well-valued with respect to the model ${\cal M}$.
Finally, we say that ${\cal H}$ is \emph{$A$-partitioned} by $[A, B]$ if ${\cal LS}({\cal H}) \cup {\cal RL}({\cal H})\subseteq A$  
 and ${\cal RS}({\cal H})\cup {\cal LL}({\cal H}) \subseteq B$. Define \emph{$B$-partitioned} analogously.

\subsection{Reduction graph}
\label{ss:redg}

In this subsection, we formally present our construction. 
We will make a reduction from \textsc{MaxCut} on cubic graphs. So, let $G$ be a cubic graph on $n$ vertices and $m$ edges. 
Intuitively, we consider an ordering of the edges of $G$, and we divide the real line into $m$ regions, with the $j$-th region holding the information about whether the $j$-th edge is in the cut-set. For this, each vertex $u$ will be related to a subset of intervals traversing all the $m$ 
regions, bringing the information about which part of the cut contains $u$.
Let $\pi_{V} = (v_1,\ldots,v_{n})$ be an ordering of $V(G)$, $\pi_{E} = (e_{1},\ldots,e_m)$ be an ordering of $E(G)$, 
and $\mathfrak{G}=(G,\pi_{V},\pi_{E})$.

 \begin{figure}[ht]\centering\captionsetup[subfigure]{justification=centering}
		\includegraphics[width=\textwidth]{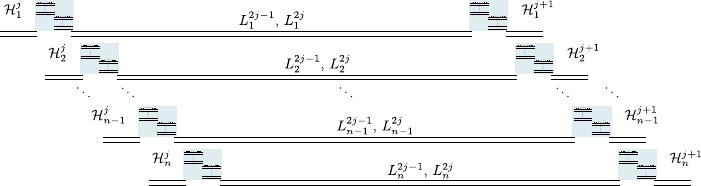}
	\caption{General structure of a region of an $(n,m)$-escalator. 
	The shaded rectangles represent the $(p,q)$-grained gadgets ${\cal H}_{i}^{j}$.}\label{fig:escalator_zoomed}
\end{figure}

We first describe the gadgets related to the vertices.
Please refer to Figure~\ref{fig:escalator_zoomed} to follow the construction.
The values of $p,q$ used next will be defined later. 
 An \emph{$(n,m)$-escalator} is an interval model ${\cal D}$ formed by $m+1$ $(p,q)$-grained gadgets for each $v_i$, denoted by ${\cal H}_i^1, \ldots, {\cal H}_i^{m+1}$, together with $2m$ \emph{link intervals}, denoted by $L_i^1,\ldots,L_i^{2m}$, such that $L_i^{2j-1}$ and $L_i^{2j}$ weakly intersect ${\cal H}_i^j$ to the right and weakly intersect ${\cal H}_i^{j+1}$ to the left.
 Additionally, all the grained gadgets are mutually disjoint. 
 More specifically, given $j\in \{1,\ldots,m+1\}$ and $i,i'\in \{1,\ldots,n\}$ with $i< i'$, the grained gadget ${\cal H}_i^j$ occurs to the left of ${\cal H}_{i'}^j$, and the grained gadget ${\cal H}_n^j$ occurs to the left of ${\cal H}_{1}^{j+1}$ for $j\in \{1,\ldots,m\}$.

\begin{figure}[ht]\centering\captionsetup[subfigure]{justification=centering}
	\includegraphics[width=\textwidth]{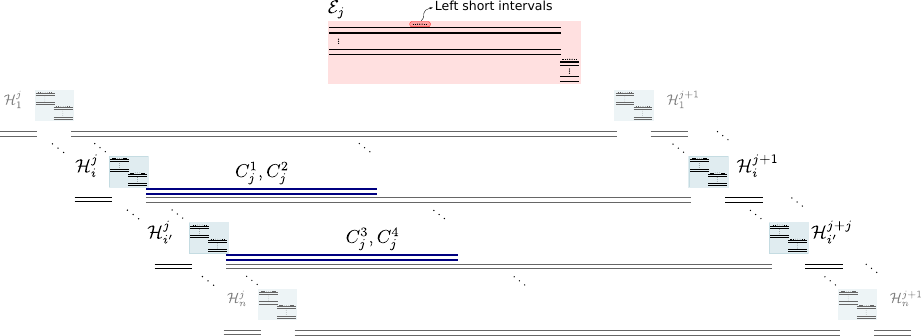}
	\caption{General structure of the constructed interval model ${\cal M}(\mathfrak{G})$ highlighting the intersections between the intervals of the $(n,m)$-escalator ${\cal D}$, the intervals of the $(p', q')$-grained gadget ${\cal E}_{j}$, and the intervals 
	$C_{j}^{1}, C_{j}^{2}, C_{j}^{3}, C_{j}^{4}$. 
    }
	\label{fig:constructed_graph}
\end{figure}

Now, we add the gadgets related to the edges. 
Please refer to  Figure~\ref{fig:constructed_graph} to follow the construction. 
The values of $p',q'$ used next will be defined later. 
For each edge $e_{j} = v_{i}v_{i'} \in E(G)$, with $i < i'$, create a $(p', q')$-grained gadget ${\cal E}_{j}$ and intervals 
$C_{j}^{1}, C_{j}^{2}, C_{j}^{3}, C_{j}^{4}$ in such a way that ${\cal E}_{j}$ is entirely contained in the $j$-th region (i.e., in the open interval between the right endpoint of 
${\cal H}_n^j$ and the left endpoint of 
${\cal H}_1^{j+1}$), $C_j^1$ and $C_j^2$ weakly intersect ${\cal H}^j_i$ to the right and weakly intersect ${\cal E}_j$ to the left, and $C_j^3$ and $C_j^4$ weakly intersect $H^j_{i'}$ to the right and strongly intersect ${\cal E}_j$ to the left. We call the intervals in $\{C^i_j\mid i\in \{1,\ldots,4\}, j\in \{1,\ldots,m\}\}$ intervals \emph{of type $C$}. Denote the constructed model by ${\cal M}(\mathfrak{G})$ (or simply by ${\cal M}$ when $\mathfrak{G}$ is clear from the context), which defines the reduction graph $\mathbb{G}_{{\cal M}(\mathfrak{G})}$.

The following straightforward lemma will be useful in the next section. 

\begin{lemma}\label{lem:parity}
Let $G$ be a graph, $\pi_{V} = (v_1,\ldots,v_{n})$ and $\pi_{E} = (e_{1},\ldots,e_m)$ be orderings of $V(G)$ and $E(G)$, respectively, and  ${\cal M}$ be the model constructed as before. 
The following holds for  every grained gadget ${\cal H}$:
\begin{itemize}
    \item ${\cal M}$ respects the structure of ${\cal H}$;
    \item The number of intervals covering ${\cal H}$ is even; and
\item The number of intervals strongly intersecting ${\cal H}$ to the left is either zero or two, and the number of intervals strongly intersecting ${\cal H}$ to the right is always zero.
\end{itemize}
\end{lemma}

Observe that Lemma~\ref{lem:parity} implies that, in order for the values $\ell$ and $r$ in Lemma~\ref{lemma:short_and_long_partition} to be odd, it suffices to choose odd values for $q$ and $q'$.

\subsection{Proof of Theorem~\ref{theo:main_result}: Maximum cut of the reduction graph}
\label{ss:prop}

Consider a cubic graph $G$ on $n$ vertices and $m=3n/2$ edges, and let $\pi_{V} = (v_1,\ldots,v_{n})$ be an ordering of $V(G)$, $\pi_{E} = (e_{1},\ldots,e_m)$ be an ordering of $E(G)$ and $\mathfrak{G}=(G,\pi_{V},\pi_{E})$. 
We now prove that 
$\mathsf{mc}(G) \geq k$ if and only if $\mathsf{mc}(\mathbb{G}_{{\cal M}(\mathfrak{G})})\geq f(G,k)$, 
where $f$ is a 
polynomial-time computable function defined at the end of this subsection.
As it is usually the case in this kind of reduction, given a cut  of $G$, 
constructing an appropriate cut of the reduction graph $\mathbb{G}_{{\cal M}(\mathfrak{G})}$ is an easy task. 
On the other hand, constructing an appropriate cut $[X,Y]$ of $G$, from a given a cut $[A,B]$ of the reduction graph $\mathbb{G}_{{\cal M}(\mathfrak{G})}$, requires that the intervals in ${\cal M}(\mathfrak{G})$ behave in a way with respect to $[A,B]$ so that $[X,Y]$ can be inferred, a task achieved by appropriately manipulating the values of $p,q,p',q'$, as 
done in Lemma~\ref{lemma:short_and_long_partition}. 
 We start by giving conditions on these values that ensure that the partitioning of the edge gadget related to an edge $e_j = v_iv_{i'}$, with $i< i'$, depends solely on the partitioning of ${\cal H}_{i'}^j$.

\begin{lemma}\label{lem:edgepartition}
Let $G$ be a cubic graph, $\pi_{V} = (v_1,\ldots,v_{n})$ and $\pi_{E} = (e_{1},\ldots,e_m)$ be orderings of $V(G)$ and $E(G)$, respectively, and  ${\cal M}(\mathfrak{G})$ be the model constructed as before, where $\mathfrak{G} = (G,\pi_V,\pi_E)$. Also, let $[A,B]$ be a maximum cut of $\mathbb{G}_{{\cal M}(\mathfrak{G})}$, and consider $e_j = v_iv_{i'}$, $i<i'$. If ${\cal M}(\mathfrak{G})$ is well-valued, and $q>4n+p'+q'+3$, then \begin{enumerate}
    \item If ${\cal H}_{i}^{j}$ is $A$-partitioned by $[A,B]$, then $\{C_{j}^{1}, C_{j}^{2}\}\subseteq B$; otherwise, $\{C_{j}^{1}, C_{j}^{2}\}\subseteq A$; and \label{item:consistence_2}
    
	\item If ${\cal H}_{i'}^{j}$ is $A$-partitioned by $[A,B]$, then $\{C_{j}^{3}, C_{j}^{4}\}\subseteq B$ and ${\cal E}_{j}$ is $A$-partitioned by $[A,B]$; otherwise, $\{C_{j}^{3}, C_{j}^{4}\}\subseteq A$ and ${\cal E}_{j}$ is $B$-partitioned by $[A,B]$. \label{item:consistence_3}
\end{enumerate}
\end{lemma}
\begin{proof}
Denote ${\cal M}(\mathfrak{G})$ by ${\cal M}$ for simplicity. 
Since ${\cal M}$ is well-valued, 
by Lemma~\ref{lemma:short_and_long_partition}, we may assume that ${\cal H}_{i}^{j}$ is $A$-partitioned by $[A,B]$, i.e., that ${\cal LS}\cup{\cal RL} \subseteq A$ and ${\cal LL}\cup {\cal RS}\subseteq B$.
We make the arguments for $C^1_j$ and it will be clear that they also hold for $C^2_j$. Observe first that all the grained gadgets covered by $C^1_j$ have a balanced number of intervals in $A$ and in $B$. More formally, from the intervals within the gadgets 
${\cal H}_\ell^j$, $i+1 \le \ell  \le n$, which are all the grained gadgets
covered by $C^1_j$, there are exactly $(n-i)(p+q)$ intervals in $A$, and $(n-i)(p+q)$ intervals in $B$.
Additionally, there are at most $2(n-i)$ link intervals intersecting $C^{1}_{j}$ to the left 
(these are the link intervals related to $v_{i''}$ for $i''>i$ in the $(j-1)$-th region, if $j>1$), 
exactly $2(n-i)$ link intervals intersecting $C^{1}_{j}$ to the right  
(these are the link intervals related to $v_{i''}$ for $i''>i$ in the $j$-th region), and exactly $2i$ link intervals covering $C^1_j$ (these are the link intervals related to $v_{i''}$ for $i''\le i$ in the $j$-th region). 
This is a total of at most $2(n-i) + 2(n-i) + 2i = 4n-2i < 4n$ link intervals.
Adding finally $C^2_j,C^3_j,C^4_j$ and the $q'$ right long intervals of ${\cal E}_j$, we get that the number of neighbors of $C^1_j$ that might be in $B$ is at most $(n-i)(p+q)+4n+q'+3$, while the number of neighbors of $C^1_j$ that are in $A$ is at least $(n-i)(p+q)+q$. 
Since $q>4n+p'+ q' +3 \ge 4n+q' +3$, we can conclude that there are more neighbours of $C^1_j$ in 
$A$ than in $B$. From  Remark~\ref{r:half}, it follows that $C^1_j \in B$.

Observe that a similar argument can be applied to $C^3_j,C^4_j$, except that we gain also $p'$ new edges from the left short intervals of ${\cal E}_j$.
That is, supposing ${\cal H}_{i'}^{j}$ is $A$-partitioned by $[A,B]$, then the number of neighbors of $C^3_j$ that might be in $B$ is at most $(n-i')(p+q)+4n+p'+q'+3$, while the number of neighbors of $C^3_j$ that are in $A$ is at least $(n-i')(p+q)+q$. 
It follows again by Remark~\ref{r:half} that $C^3_j,C^4_j$ are in $B$, since $q>4n+p'+ q' +3$.

Finally, suppose that ${\cal H}_{i'}^{j}$ is $A$-partitioned by $[A,B]$, in which case, from the previous paragraph, we get that $\{C^3_j,C^4_j\}\subseteq B$.
Suppose that ${\cal E}_j$ is $B$-partitioned. 
Then consider the cut $[A',B']$ obtained by switching the intervals of ${\cal E}_j$ of side; formally, in which every interval $I \in {\cal M}\setminus {\cal E}_j$ is in $A'$ if and only if $I \in A$ and every interval $I \in {\cal E}_j$ is in $A'$ if and only if $I \in B$. 
Clearly, the number of cut-edges having both endpoints in ${\cal E}_j$ is the same in both the cuts $[A,B]$ and $[A',B']$. Since
$\lvert A \cap {\cal E}_j \rvert$ = $\lvert B \cap {\cal E}_j \rvert$ = $\lvert A' \cap {\cal E}_j \rvert$ = $\lvert B' \cap {\cal E}_j \rvert$, and every interval other than $C^1_j,C^2_j,C^3_j$ and $C^4_j$ that intersects the gadget ${\cal E}_j$ has a cover intersection with it, the number of cut-edges in $[A, B]$ differ from that of $[A',B']$ only by the number of cut-edges between $C^1_j,C^2_j,C^3_j,C^4_j$ and ${\cal E}_j$. 
Since $C^3_j,C^4_j \in B$, the $2p'$ edges between these two intervals and the intervals in ${\cal LS}({\cal E}_j)$ are cut-edges in $[A',B']$ but not in $[A,B]$.
Meanwhile, the edges between $C^1_j,C^2_j,C^3_j,C^4_j$ and ${\cal E}_j$ that are cut-edges in $[A, B]$ but not in $[A',B']$ must be from among the $4q'$ edges between $C^1_j,C^2_j,C^3_j,C^4_j$ and ${\cal LL}({\cal E}_j)$. 
Thus $[A',B']$ has at least $2p' - 4q'$ edges more than $[A,B]$. Since ${\cal M}$ is well-valued, we have $2p' > 4q'$, implying that $[A',B']$ is a cut of size larger than $[A, B]$, which is a contradiction.
\end{proof}

After ensuring that each grained gadget behaves well individually, we also need to ensure that ${\cal H}^1_i$ can be used to decide in which part of $[X,Y]$ we should put $v_i$, and for this it is necessary that all gadgets related to $v_i$ agree with one another. In other words, for each $v_i$, we want that the behaviour 
of the first gadget ${\cal H}^1_i$ influence the behaviour of the subsequent gadgets ${\cal H}^2_i, \ldots, {\cal H}^{m+1}_i$, as well as the behaviour of the gadgets related to edges incident to $v_i$. 
Given $v_i\in V(G)$ and a cut $[A,B]$ of $\mathbb{G}_{{\cal M}(\mathfrak{G})}$, we say that \emph{the gadgets of  $v_i$ alternate in $[A,B]$} if, for every $j\in \{1,\ldots,m\}$, we get that ${\cal H}^j_i$ is $A$-partitioned if and only if ${\cal H}^{j+1}_i$ is $B$-partitioned, 
while $L^{2j-1}_i,L^{2j}_i$ 
are opposite to the right long intervals of ${\cal H}^{j}_i$. Also, we say that $[A,B]$ is \emph{alternating partitioned} if the gadgets of $v_i$ alternate in $[A,B]$, for every $v_i\in V(G)$. We add a further condition on the values of $p,q,p',q'$ in order to ensure that every maximum cut is alternating partitioned. After this, we use the good behaviour of the constructed model in order to relate the sizes of the maximum cuts  in $G$ and in $\mathbb{G}_{{\cal M}(\mathfrak{G})}$.

\begin{lemma}\label{l:ABcomunica}
Let $G$ be a cubic graph, $\pi_{V} = (v_1,\ldots,v_{n})$ and $\pi_{E} = (e_{1},\ldots,e_m)$ be orderings of $V(G)$ and $E(G)$, respectively, and  ${\cal M}(\mathfrak{G})$ be the model constructed as before, where $\mathfrak{G} = (G,\pi_V,\pi_E)$. Also, let $[A,B]$ be a maximum cut of $\mathbb{G}_{{\cal M}(\mathfrak{G})}$. If ${\cal M}(\mathfrak{G})$ is well-valued, 
$q >4n+p'+q' + 3$,
and 
$q>3(2n^2+2n+q'+2)$, 
then 
$[A,B]$ is alternating partitioned. \end{lemma}
\begin{proof}
By hypothesis, the conditions of Lemmas~\ref{lemma:short_and_long_partition} and~\ref{lem:edgepartition} are satisfied. 
Thus, we can suppose that the obtained properties of those lemmas hold. 
Denote ${\cal M}(\mathfrak{G})$ by ${\cal M}$ for simplicity, and let ${\cal M}_i$ be the family of all the intervals related to vertex $v_i$; more specifically, it contains every interval in some grained gadget ${\cal H}_i^j$, $j\in \{1,\ldots,m+1\}$, every link interval $L_i^j$, $j\in \{1,\ldots, 2m\}$, every interval of type $C$ that intersects ${\cal H}^j_i$ to the right (this happens if $e_j$ has $v_i$ as endpoint), and every interval in ${\cal E}_j$ for $e_j$ incident to $v_i$. In what follows, we count the number $f_i$ of edges of the cut incident to some interval in ${\cal M}_i$ and argue that, if the gadgets of $v_i$ do not alternate in $[A,B]$, then we can obtain a bigger cut 
by switching the side of some intervals, thus getting a contradiction. 

Denote by ${\cal \overline{M}}_i$ the set of intervals ${\cal M}\setminus {\cal M}_i$, and by $\Lambda$
the set of all link intervals. 

In what follows, there are some values that must be added to $f_i$ but remain the same in every maximum cut of $\mathbb{G}_{{\cal M}(\mathfrak{G})}$, independently of how ${\cal M}_i$ is partitioned; we call these values \emph{irrelevant} and do not add them to $f_i$. 
For instance, recall that every $(x,y)$-grained gadget has exactly $x+y$ intervals in $A$ and $x+y$ in $B$. Thus, because of Lemmas~\ref{lemma:short_and_long_partition} and~\ref{lem:edgepartition}, the number of edges of the cut between grained gadgets and intervals that cover them is irrelevant. In what follows, we count the other possible edges.

First, consider $j\in \{1,\ldots,m\}$; we want to count the maximum number of edges of the cut incident to $L^{2j}_i$ (which holds analogously for $L^{2j-1}_i$). Denote by $\ell^j_A$ the number of intervals in 
${\cal \overline{M}}_i\cap \Lambda \cap A$ 
that intersect $L^{2j}_i$; define $\ell^j_B$ similarly. Observe that $\ell^j_A+\ell^j_B < 4n$
since it includes at most $2(n-i)$ link intervals in the $j$-th region, plus at most $2(n-i)$ link intervals of the $(j-1)$-th region, and at most $2(i-1)$ link intervals of the $(j+1)$-th region.
Additionally, let $a_j$ be equal to 1 if $L^{2j}_i$ is opposite to the right long intervals of ${\cal H}^j_i$, and 0 otherwise; similarly, let $b_j$ be equal to 1 if $L^{2j}_i$ is opposite to the left long intervals of ${\cal H}^{j+1}_i$, and 0 otherwise. Because $L^{2j}_i$ might also be opposite to $C^1_j,\ldots,C^4_j$
and it is possible that the edge between $L^{2j}_i$ and $L^{2j-1}_i$ is also a cut edge, observe that the relevant number of edges of the cut incident to $L^{2j}_i$ is at most $q(a_j+b_j)+\ell^j_A+\ell^j_B+5$. 
Note that $L^{2j}_i$ covers the gadgets ${\cal E}_j$ and also every ${\cal H}^{j'}_{i'}$ with which it has an intersection except ${\cal H}^j_i$ and  ${\cal H}^{j+1}_i$, and hence the number of cut-edges between $L^{2j}_i$  and intervals in these gadgets is irrelevant.

Now, let $e_j$ be an edge incident to $v_i$ and let $v_{i'}$ be the other endpoint of $e_j$ (here $i'$ might be smaller than $i$). We apply Lemma~\ref{lem:edgepartition} in order to count the edges incident to ${\cal E}_j\cup \{C^1_j,\ldots,C^4_j\}$; observe that all these intervals are in ${\cal M}_i$. First observe that, since ${\cal E}_j$ is always partitioned according to $C^3_j,C^4_j$, we have an irrelevant value of $2p'$, namely the edges between $C^3_j,C^4_j$ and the left short intervals of ${\cal E}_j$. 
Now, suppose, without loss of generality, that $\{C^1_j,C^2_j\} \subseteq A$. If  $\{C^3_j,C^4_j\}\subseteq A$, then there are no relevant edges to be added; otherwise, we get $2q'+4$ edges, those between $C^1_j,C^2_j$ and $C^3_j,C^4_j$, and between $C^1_j,C^2_j$ and the left long intervals of ${\cal E}_j$. Finally, observe that the edges between $\{C^1_j,\ldots,C^4_j\}$ and ${\cal H}^j_i$ are irrelevant because of Lemma~\ref{lem:edgepartition} and the fact that $C^1_j$, $C^2_j$ cover ${\cal H}^j_{i'}$ (where $e_j$ is the edge $v_iv_{i'}$), and that the edges 
between $\{C^1_j,\ldots,C^4_j\}$
and the link intervals have been counted previously. 
Note that the number of cut edges between two intervals in ${\cal E}_j$ and the number of cut edges between intervals in ${\cal E}_j$ and link intervals are both irrelevant.
Note also that every gadget ${\cal H}^j_i$, for each $j \in \{1, 2, \ldots , m+1\}$, is covered by every interval from $\overline{{\cal M}}_i$ that it intersects, and hence the number of cut edges between vertices in ${\cal H}^j_i$ and vertices in $\overline{{\cal M}}_i$ is irrelevant. Also, the number of cut edges between vertices in ${\cal H}^j_i$ is irrelevant, and the cut edges having one endpoint in ${\cal H}^j_i$ and other endpoint in ${\cal M}_i \setminus {\cal H}^j_i$ have already been counted.

In order to put everything together, let $e_{j_1}$, $e_{j_2}$, $e_{j_3}$ be all the edges incident to $v_i$, and for each $h\in \{1,2,3\}$, write $e_{j_h} = v_iv_{i_h}$ (note that here $i$ is not necessarily smaller than $i_h$).
For each $h\in \{1,2,3\}$, let  $c_{h}$ be equal to 1 if ${\cal H}^j_i$ and ${\cal H}^j_{i_h}$ are partitioned differently, and 0 otherwise. We then get that:

\begin{equation}f_i \le 2\sum_{j=1}^m (q(a_j+b_j) + \ell^j_A+\ell^j_B + 5) + \sum_{h=1}^3 c_{h}(2q'+4).\label{eq:fi}\end{equation}

If $L^{2j}_i$ is on the same side as the  right long intervals of ${\cal H}^j_i$ and the left long intervals of ${\cal H}^{j+1}_i$, we can increase $f_i$ simply by switching the side of $L^{2j}_i$. 
Indeed, in this case we would lose at most $\max\{\ell^j_A,\ell^j_B\}+5 < 4n+5$
edges, while gaining $2q$, a positive exchange since $2q > 8n > 4n+5$ considering $n>1$. 
Observe that this implies $a_j+b_j\ge 1$. Note also that this type of argument can be always applied, and that it can be applied also for $L^{2j-1}_i$. Hence, whenever in what follows we switch side of the intervals in some vertex gadget, we can suppose that this property still holds, i.e. that $L^{2j}_i$ and $L^{2j-1}_i$ are always opposite to the left long intervals of ${\cal H}^j_i$.

Consider now $j$ to be minimum such that ${\cal H}^j_i$ and ${\cal H}^{j+1}_i$ are partitioned in the same way, say they are both $A$-partitioned. Note that this implies that $a_j+ b_j = 1$, since the right long intervals of ${\cal H}^j_i$ are in $A$, while the left long intervals of ${\cal H}^{j+1}_i$ are in $B$. We want to switch sides of ${\cal H}^{j+1}_i$, but in order to ensure an increase in the size of the cut, we need to also switch subsequent grained gadgets in case they were alternating. 
For this, let $j' > j$ be minimum such that ${\cal H}^{j'+1}_i$ and ${\cal H}^{j'}_i$ are either both $A$-partitioned or both $B$-partitioned; if it does not exist, let $j'=m+1$. 
For each $h\in \{j+1,\ldots, j'\}$, we switch sides of ${\cal H}^h_i$, and put $L^{2h-1}_i,L^{2h}_i$ in the side opposite to the right long intervals of ${\cal H}^h_i$. Also switch the intervals of type $C$ and intervals in edge gadgets appropriately; i.e., in a way that Lemma~\ref{lem:edgepartition} continues to hold. 
We prove that we gain at least $2q$ edges, 
while losing at most $8m(n+1) + 6(q'+2) = 6(2n^2+2n+q'+2)$ cut edges (recall that $m = 3n/2$); 
the result thus follows since $q>3(2n^2+2n+q'+2)$. 

Observe that, by previous arguments, we have that, for every $h \in \{j, \ldots, j'\}$, the link intervals $L^{2h-1}_i,L^{2h}_i$ are in $B$ if and only if ${\cal H}^h_i$ is $A$-partitioned. In particular, since ${\cal H}^j_i$ is $A$-partitioned, $L^{2j-1}_i$ and $L^{2j}_i$ are in $B$. Additionally, because of the switch we now know that the left long intervals of ${\cal H}^{j+1}_i$ are in $A$. 
This implies that we gain at least $2q$ edges. 
Now, we count our losses. Concerning intervals $L^{2j-1}_i$ and $L^{2j}_i$, we lose at most $2(\ell_B^j+4) \le 8n+8$ cut edges, 
namely the edges between these intervals and link intervals or intervals of type $C$. As for the intervals $L^{2h-1}_i,L^{2h}_i$ for $h\in \{j+1,\ldots,j'\}$, by the definition of $j'$ we know that we lose at most $2(\max\{\ell_A^h,\ell_B^h\}+4) \le 8n+8$ cut edges,
while the number of edges of the cut between them and the vertex grained gadgets can only increase. Hence, concerning the link intervals in ${\cal M}_i$, in total we lose at most $8m(n+1) = 12(n^2+n)$ cut edges.
Additionally, 
observe the upper bound given by~(\ref{eq:fi}) to see that,
in the worst case scenario, we have $\{j_1,j_2,j_3\}\subseteq \{j+1,\ldots,j'\}$ and all the values $c_{h}$ were previously equal to 1 
and are now equal to 0; 
this leads to a possible loss of at most $6(q'+2)$ edges, as we wanted to show.
\end{proof}

Now, if $[A,B]$ is an {alternating partitioned} maximum cut of $\mathbb{G}_{{\cal M}(\mathfrak{G})}$, and ${\cal M}(\mathfrak{G})$ obeys the conditions in the statement of Lemma~\ref{lem:edgepartition}, we let $\Phi(A,B) = [X,Y]$ be the cut of $G$ such that, for each vertex $v_{i} \in V(G)$, we have $v_{i} \in X$ if and only if ${\cal H}_{i}^{1}$ is $A$-partitioned by $[A,B]$. 
Note that $[X,Y]$ is well-defined (i.e., $\Phi$ is a function). Additionally, given a cut $[X,Y]$ of $G$, there is a unique alternating partitioned cut $[A,B]$ of $\mathbb{G}_{{\cal M}(\mathfrak{G})}$
obeying the conditions  of Lemma~\ref{lem:edgepartition}, such that $[X,Y]=\Phi(A,B)$ (i.e., $\Phi$ is one-to-one and onto). 
Therefore, it remains to relate the sizes of these cut-sets. 
Basically we can use the good behaviour of the maximum cuts in $\mathbb{G}_{{\cal M}(\mathfrak{G})}$ to prove that the size of $[A,B]$ grows as a function of the size of $\Phi(A,B)$.

\begin{lemma}\label{lem:function}
Suppose that all the conditions in Lemmas~\ref{lemma:short_and_long_partition}-\ref{l:ABcomunica} hold, and that $q'\ge 13n^2$. Let $\Phi(A,B) = [X,Y]$, and $k$ be a positive integer. Then (below, $\mathbb{G}$ denotes $\mathbb{G}_{{\cal M}(\mathfrak{G})}$)
\[\lvert E_G(X,Y)\rvert \ge k\text{ if and only if }\lvert E_{\mathbb{G}}(A,B)\rvert \ge \gamma + (2q'+4)k,\]
where $\gamma $ is a well-defined polynomial-time computable function on $G, \pi_{V},p,q,p',q'$ (i.e., does not depend on $[A,B]$).
\end{lemma}
\begin{proof}
We use the same notation as before and count the number of edges in $E_{\mathbb{G}}(A,B)$. We will count the number of edges of the cut-set separately in the following groups: \\
\begin{itemize}
\item among intervals of a vertex/edge grained-gadget;
 \item between intervals of a vertex grained-gadget and link intervals;
\item between intervals of an edge grained-gadget and other intervals;
\item among intervals of type $C$;
\item among link intervals;
\item between link intervals and intervals of type $C$; and
\item between intervals of a vertex grained-gadget and intervals of type $C$.
\end{itemize}

First, we compute the number of edges of the cut-set within a given $(x,y)$-grained gadget. By Lemma~\ref{lemma:short_and_long_partition}, we get that this is exactly $y^2+2xy$. Since there are $(m+1)n$ $(p,q)$-grained gadgets (the ones related to the vertices), and $m$ $(p',q')$-grained gadgets (the edge ones), we get a total of:

\[\beta_1 = n(m+1)(q^2+2pq)+ m((q')^{2}+2p'q').\]

Now, we count the number of edges of the cut-set between a given vertex grained gadget ${\cal H} = {\cal H}^j_i$ and link intervals; again, denote the set of link intervals by $\Lambda$.
If an interval $I$ covers ${\cal H}$, then there are exactly $p+q$ edges between $I$ and ${\cal H}$, since there are these many intervals of ${\cal H}$ in each of $A$ and $B$. And if $I$ intersects ${\cal H}$ either to the left or to the right, then there are exactly $q$ edges between $I$ and ${\cal H}$, since ${\cal M}$ is alternating partitioned (i.e., $I$ is opposite to the corresponding long intervals of ${\cal H}$). 
It remains to count how many of each type of intervals there are. If $j\in \{2,\ldots,m\}$, then there are exactly $2n-2$ intervals covering ${\cal H}$, as well as $2$ intervals intersecting ${\cal H}$ to the left, and $2$ to the right; this gives a total of $(2n-2)(p+q)+4q = 2n(p+q)+2(q-p)$ edges between ${\cal H}$ and $\Lambda$.
If $j=1$, then there are $2(i-1)$ intervals covering ${\cal H}$, and $2$ intervals intersecting ${\cal H}$ to the right, thus giving a total of $2(i-1)(p+q)+2q$. Finally, if $j=m+1$, then there are $2(n-i)$ intervals covering ${\cal H}$, and $2$ intervals intersecting ${\cal H}$ to the left, giving a total of $2(n-i)(p+q)+2q$. Summing up, we get:

\[\begin{array}{rl}
\beta_2 = & \sum_{j=2}^m\sum_{i=1}^n [2n(p+q)+2(q-p)] +
\\ & \sum_{i=1}^n [2(i-1)(p+q)+2q+2(n-i)(p+q)+2q]\\
 = & 2(m-1)n[n(p+q)+(q-p)] + 2n[(n-1)(p+q)+2q]\\
 = & 2n [ (m-1)n(p+q) + (m-1)(q-p) + (n-1)(p+q) + 2q]\\
= & 2mn[n(p+q)+q-p].
\end{array}\]

We count now the number of edges of the cut-set between a given edge gadget ${\cal E}_j$ and an interval~$I$ intersecting it, and among intervals of type $C$. As before, if $I$ covers ${\cal E}_j$, then there are exactly $(p'+q')$ edges between $I$ and ${\cal E}_j$ in the cut. If $I$ strongly intersects ${\cal E}_j$ to the left, then $I\in \{C^3_j,C^4_j\}$ and by Lemma~\ref{lem:edgepartition} we get that this amounts to $p'$. Finally, if $I$ weakly intersects ${\cal E}_j$ to the left, then this amounts to $q'$, if $e_j$ is in the cut-set, or to $0$, otherwise. As for the number of edges between intervals of type $C$, by Lemma~\ref{lem:edgepartition} one can see that this is equal to $4\lvert E_G(X,Y)\rvert $. Summing up, we get:
\[2nm(p'+q') + 2p'm + (2q'+4)\lvert E_G(X,Y)\rvert.\]
Denote the value $2nm(p'+q')+2p'm$ by $\beta_3$, and note that this is independent of $[A,B]$. 

Let us now count the number of edges of the cut-set among link intervals. For this, denote by ${\cal L}^j$ the set of link intervals in the $j$-th region, i.e., ${\cal L}^j = \{L^{2j}_i,L^{2j-1}_i\mid i\in \{1,\ldots,n\}\}$. 
Also, denote by $V^j_A$ the set of indices $i\in \{1,\ldots,n\}$ such that $\{L^{2j-1}_{i},L^{2j}_{i}\}\subseteq A$; define $V^j_B$ analogously and let $a = \lvert V^j_A\rvert$ and $b = \lvert V^j_B\rvert$. 
We count the number of edges of the cut between intervals of ${\cal L}^j$, for every $j\in \{1,\ldots,m\}$, and between intervals of ${\cal L}^{j}$ and intervals of ${\cal L}^{j+1}$, for every $j\in \{1,\ldots,m-1\}$, and then we sum up. So consider a region $j\in \{1,\ldots,m\}$, and observe that, because $[A,B]$ is alternating partitioned, we get that either $j$ is odd and $V^j_A$ contains exactly the indices of the vertices within $Y$,
while $V^j_B$ contains the indices of the vertices within $X$,
or $j$ is even and the reverse occurs. More formally: if $j$ is odd, then 
$V^j_A = \{i\in \{1,\ldots,n\}\mid v_i\in Y\}$
and $V^j_B = \{i\in \{1,\ldots,n\}\mid v_i\in X\}$;
and if $j$ is even, then 
$V^j_A = \{i\in \{1,\ldots,n\}\mid v_i\in X\}$ 
and $V^j_B = \{i\in \{1,\ldots,n\}\mid v_i\in Y\}$.
In either case, since for each index in $V^j_A$ (resp. $V^j_B$), there is a pair of intervals in ${\cal L}^j\cap A$ (resp. ${\cal L}^j\cap B$), we get that the number of edges of the cut between intervals of ${\cal L}^j$ is equal to $4\lvert X\rvert\lvert Y\rvert = 4ab$. 
Now, suppose $j\in \{1,\ldots,m-1\}$; we count the edges of the cut between ${\cal L}^j$ and ${\cal L}^{j+1}$. Again because $[A,B]$ is alternating partitioned, we know that if $V^j_A = \{i_1,\ldots,i_a\}$, then $V^{j+1}_B = V^j_A$, while $V^{j+1}_A = V^j_B = \{1,\ldots,n\}\setminus V^j_A$. Supposing $i_1 < \cdots < i_a$, this implies that there are exactly $4$ edges between $\{L^{2j+1}_{i_{a'}},L^{2j+2}_{i_{a'}}\}$ and $\{L^{2j-1}_{i_{a''}},L^{2j}_{i_{a''}}\}$ for each $a',a''\in \{1,\ldots,a\}$ with $a'<a''$. Summing up we get that there are $4\sum_{a'=1}^a (a-a') = 4\frac{a(a-1)}{2} = 2a(a-1)$ edges between ${\cal L}^j\cap A$ and ${\cal L}^{j+1}\cap B$. Analogously we can conclude that there are $2b(b-1)$ edges between ${\cal L}^j\cap B$ and ${\cal L}^{j+1}\cap A$. Summing up with the previous $4ab$, for every $j\in \{1,\ldots,m-1\}$, we get $2a^2-2a+2b^2-2b+4ab = 2[(a+b)^2-(a+b)]$ edges of the cut incident to ${\cal L}^j$ 
 minus the number of edges of the cut between ${\cal L}^j$ and ${\cal L}^{j-1}$, as these get counted in ${\cal L}^{j-1}$. 
Recall that $a+b = \lvert X\rvert +\lvert Y\rvert = n$ to see that this gives us $2n(n-1)$ edges. Finally, summing up for all $j\in\{1,\ldots,m-1\}$ and summing also the edges between link intervals in ${\cal L}^m$, we get that the number of edges of the cut incident to link intervals is equal to:
\[\sum_{j=1}^{m-1}2n(n-1) + 4\lvert X\rvert \lvert Y\rvert = n(n-1)(3n-2) + 4\lvert X\rvert \lvert Y\rvert\]
Observe that $4(n-1)\le 4\lvert X\rvert \lvert Y\rvert \le n^2$, and denote the value $n(n-1)(3n-2)$  by $\beta_4$.

Now, observe that it remains to count the number of edges of the cut-set between link intervals and intervals of type $C$, and between intervals of type $C$ and vertex grained gadgets. We start with the latter. Given an  edge $e_j = v_iv_{i'}$, with $i < i'$, there are exactly $n-i$ vertex grained gadgets covered by $C^1_j,C^2_j$, and $n-i'$ vertex grained gadgets covered by $C^3_j,C^4_j$. Together with the $q$ edges between each of these intervals of type $C$ and the corresponding vertex gadgets (namely, ${\cal H}^j_i$ and ${\cal H}^j_{i'}$), we get a total of $2(n-i)(p+q)+2(n-i')(p+q)+4q$. Even though we cannot give a precise value below, observe that this value can be exactly computed during the construction. The upper bound is given just to make it explicit that this is a polynomial function. Also, below, for $e_j = v_iv_{i'}$, the value $\ell_j$ denotes $i$ and $r_j$ denotes $i'$.

\[\begin{array}{rl}
\beta_5 = & \sum_{j=1}^m [2(n-r_j)(p+q)+2(n-\ell_j)(p+q)+4q]\\
= & \sum_{j=1}^m [4n(p+q)-2(r_j+\ell_j)(p+q) + 4q] \\ \le & 4m[n(p+q)+q].
\end{array}\]

Finally, we count the number of edges of the cut between link intervals and intervals of type $C$. This is the only part of the counting that will not be exact. Again, consider an edge $e_j = v_iv_{i'}$, and first consider the interval $C^1_j$; we will see that the arguments hold for $C^2_j$, and that analogous arguments hold for $C^3_j,C^4_j$. Observe that $C^1_j$ intersects exactly the following link intervals: $L^{2j-1}_{i''}$ and $L^{2j}_{i''}$ for every $i''\in \{1,\ldots,n\}$; and $L^{2j-2}_{i''}$ and $L^{2j-3}_{i''}$ for every $i''\in \{i+1,\ldots,n\}$. 
This is a total of less than $4n$ link intervals. Because an analogous argument can be applied to $C^2_j,C^3_j,C^4_j$, we get a total of $16n$ possible edges in the cut-set, for each value of $j$, totalling $16nm = 24n^2$. 

Let $\beta = \sum_{i=1}^5\beta_i$, and $\gamma = \beta+4(n-1)$. We now prove that $\lvert E_G(X,Y)\rvert \ge k\text{ if and only if }\lvert E_{\mathbb{G}}(A,B)\rvert \ge \gamma + (2q'+4)k$. 
We have proved that:
$$\begin{array}{rl}
     \overbrace{\beta + 4(n-1)}^{\gamma} +(2q'+4)\lvert E_G(X,Y)\rvert\le &  \lvert E_{\mathbb{G}}(A,B)\rvert \\
\le & \beta+25n^2+(2q'+4)\lvert E_G(X,Y)\rvert.
\end{array}$$
If $\lvert E_G(X,Y)\rvert\ge k$, then by the first inequality we have that $\lvert E_{\mathbb{G}}(A,B)\rvert\ge \beta+4(n-1)+(2q'+4)k$. 
On the other hand, if $\lvert E_{\mathbb{G}}(A,B)\rvert \ge \beta+4(n-1) + (2q'+4)k$, then by the second inequality we have that $\lvert E_G(X,Y)\rvert \ge k - \frac{25n^2-4(n-1)}{2q'+4} \ge k - \frac{26n^2}{2q'+4}$.
Since $q'\ge 13n^2$, we get that 
$\lvert E_G(X,Y)\rvert >k-1$.
\end{proof}

To finish the proof that the reduction works, we simply need to choose appropriate values for $p,q,p',q'$. 
Recall all necessary conditions:
\begin{itemize}
    \item\label{condition1} For each $(x,y)$-grained gadget ${\cal H}$ in ${\cal M}$, let $t$ be the number of intervals in ${\cal M}\setminus{\cal H}$ intersecting ${\cal H}$, $\ell$ be the number of intervals in ${\cal M}\setminus{\cal LS}$ intersecting the left short intervals, and $r$ be the number of intervals in ${\cal M}\setminus{\cal RS}$ intersecting the right short intervals. Then we want that $\ell$ and $r$ are both odd, 
    and that  $y > t (x/y -1)$
    and $x > t + 2y$ (from Lemma~\ref{lemma:short_and_long_partition});
\item $q>4n+p'+q'+3$ (from Lemma~\ref{lem:edgepartition});
\item $q>3(2n^2+2n+q'+2)$ (from Lemma~\ref{l:ABcomunica}); and
    \item $q'\ge 13n^2$ (from Lemma~\ref{lem:function}).
\end{itemize}

By Lemma~\ref{lem:parity}, we know that in order for the values $r,\ell$ in the first item to be odd, it suffices to choose $q,q'$ to be odd. Observe that $n\ge 4$ since $G$ is a cubic graph. For a given edge gadget ${\cal E}_j$, we know that there are exactly $2n+4$ intervals in ${\cal M}\setminus {\cal E}_j$ intersecting it, namely the link intervals and intervals of type $C$ in the $j$-th region. 
We could just choose $q'\in \{13n^2,13n^2 + 1\}$ such that $q'$ is odd and $p' = 26n^2 +2n+7$. 
In this case, we have $p' > t+2q'$
and $q' \ge 13n^2 > (\frac{26n^2 +2n+7}{13n^2} - 1) (2n + 4) \ge (\frac{p'}{q'} - 1) t$,  
since for edge gadgets $t = 2n + 4$.
Similarly, we choose $q\in \{42n^2 + 3n + 10,42n^2 + 3n + 11\}$ such that $q$ is odd and $p = 84n^2 +8n+29$.
We now have $p > t+2q$ 
and $q \ge 42n^2 + 3n + 10 > (\frac{84n^2 + 8n + 29}{42n^2 + 3n + 10} - 1) (2n + 6) \ge (\frac{p}{q} - 1) t$,
since for vertex gadgets $t \le 2n + 6$.

To finish the proof of Theorem~\ref{theo:main_result}, it remains to prove that the interval count of our reduction graph is exactly four, which is done in the next subsection.

\subsection{Proof of Theorem~\ref{theo:main_result}: Bounding the interval count}
\label{ss:boun}

Consider a cubic graph $G$ on $n$ vertices and $m=3n/2$
edges, and orderings $\pi_V,\pi_E$ of the vertex set and edge set of $G$.
Denote the triple $(G,\pi_V,\pi_E)$ by $\mathfrak{G}$. 
First, we want to prove that the interval count of our constructed interval model ${\cal M}(\mathfrak{G})$ 
is at most~4. But observe that the construction of ${\cal M}(\mathfrak{G})$ is actually not unique, since the intervals are not uniquely defined; e.g., given such a model, one can obtain a model satisfying the same properties simply by adding $\epsilon>0$ to all points defining the intervals. In what follows, we provide a construction of a uniquely defined interval model related to $\mathfrak{G}$ that satisfies the desired conditions and has interval count~4. 

Consider our constructed interval model ${\cal M}(\mathfrak{G})$, and for each $j\in \{1,\ldots,m\}$,  denote by ${\cal S}_j$ the set of intervals related to the $j$-th region, i.e., ${\cal S}_j = {\cal E}_j\cup \bigcup_{\ell=1}^4 C_{j}^{\ell}\cup \bigcup_{i=1}^n({\cal H}_{i}^{j}\cup \{L^{2j}_i\cup L^{2j-1}_i\})$.
We show how to accommodate ${\cal S}_1$ within the closed interval $[0,6n-2]$ in such a way that the same pattern can be adopted in the subsequent regions of ${\cal M}(\mathfrak{G})$ too, each time starting at multiples of $4n$. More specifically, letting $t = 4n$, we will accommodate ${\cal S}_j$ within $[t\cdot (j-1),6n-2+ t\cdot (j-1)]$.  
Assume $e_1 = v_iv_{i'}$, with $i<i'$. Below, we describe exactly which closed interval of the line corresponds to each interval $I\in {\cal S}_1$.

\begin{itemize}
\item For each $i\in \{1,\ldots, n\}$, the left long intervals of ${\cal H}_i^1$ are equal to $[2i-2,2i-\frac{3}{2}]$
and the left short intervals are any choice of $p$ distinct points within the open interval $(2i-2,2i-\frac{3}{2})$,
whereas the right long intervals of ${\cal H}_i^1$ are equal to $[2i-\frac{3}{2},2i-1]$
and the right short intervals are any choice of $p$ distinct points within the open interval $(2i-\frac{3}{2},2i-1)$. 
Note that open intervals are used to locate the closed intervals of length zero, but that the short intervals themselves are not open. 
\item $C^1_1$ and $C^2_1$ are equal to $[2i-1,2i+2n-2]$.
\item $C^3_1$ and $C^4_1$ are equal to $[2i'-1, 2i'+2n-2]$.
\item The left long intervals of ${\cal E}_1$ are equal to $[2n,4n-1]$.
\item The left short intervals of ${\cal E}_1$ are any choice of $p'$ distinct points in the open interval $(2i+2n-2,2i'+2n-2)$.
Again, the open interval is used just to locate the closed intervals of length zero.
\item The right long intervals of ${\cal E}_1$ are equal to $[4n-1,4n-\frac{1}{2}]$
and the right short intervals are any choice of $p'$ distinct points within the corresponding open interval. 

\item For each $i\in \{1,\ldots,n\}$, intervals $L^1_i,L^2_i$ are equal to $[2i-1,4n+2(i-1)]$.\end{itemize}

\definecolor{myblue}{RGB}{0,0,255}
\definecolor{mygreen}{RGB}{31,120,68}
\definecolor{mygrey}{RGB}{156, 156, 156}
\definecolor{darkblue}{RGB}{0, 0, 153}

\def\epsilon{0.02} \def\n{4}
\def\m{1}
\def\h{3}
\def\hprime{4}
\def\t{4*\n}

\begin{figure}[thb]
  \begin{center}
      \resizebox{\textwidth}{!}{
        \begin{tikzpicture}[scale=1.0]
        
          \pgfsetlinewidth{1pt}
          \tikzset{thinScale/.style={dash pattern=on 1pt off 2pt, line width=0.5\pgflinewidth, mygrey}}
          \tikzset{thickScale/.style={dash pattern=on 2.5pt off 3pt, line width=2.0pt}}
          \tikzset{closeInterval/.style={{|[scale=0.8,width=6.8,line width=1.0pt, sep=-0.50pt]}-{|[scale=0.8,width=6.8,line width=1.0pt,sep=-0.50pt]}, line width=2pt}}
          \tikzset{shortInterval/.style={dot diameter=1.4pt, dot spacing=1.9pt, dots}}

\foreach \i in {1, ..., \n} {
            \foreach \j in {0, ..., \m}{
              \draw[thinScale] (2*\i-2 + \t*\j,0)--(2*\i-2 + \t*\j,-\n - 1);
              \draw[thinScale] (2*\i-1.5 + \t*\j,0)--(2*\i-1.5 + \t*\j,-\n - 1);
              \draw[thinScale] (2*\i-1 + \t*\j,0)--(2*\i-1 + \t*\j,-\n - 1);
            }
          }

          \draw[thinScale] (2*\n,0)--(2*\n,-\n - 1) (2*\h + 2*\n-2,0)--(2*\h + 2*\n-2,-\n - 1) (2*\hprime + 2*\n-2,0)--(2*\hprime + 2*\n-2,-\n - 1) (4*\n-1,0)--(4*\n-1,-\n - 1) (4*\n-0.5,0)--(4*\n-0.5,-\n - 1);
\draw[thickScale, mygreen] (\t,0)--(\t,-5);

          \foreach \i in {1, ..., \n}{
            \foreach \j in {0, ..., \m}{

\draw[shortInterval] (2*\i-2+\epsilon + \t*\j,-\i+0.1)--(2*\i-1.5 + \t*\j-\epsilon,-\i+0.1); 

\draw[shortInterval] (2*\i-1.5+\epsilon + \t*\j,-\i-0.1)--(2*\i-1 + \t*\j-\epsilon,-\i-0.1); 

\draw[closeInterval,red] (2*\i-2 + \t*\j,-\i)--(2*\i-1.5 + \t*\j,-\i);

\draw[closeInterval,red] (2*\i-1.5 + \t*\j,-\i-0.2)--(2*\i-1 + \t*\j,-\i-0.2); 
            }

\draw[closeInterval,orange] (2*\i-1,-\i-0.6)--({2*(\i-1)+ 4*\n},-\i-0.6); 

          }

\draw[closeInterval,myblue] (2*\h-1,-\h-0.4)--(2*\h + 2*\n-2,-\h-0.4);

\draw[closeInterval,myblue] (2*\hprime-1, -\hprime-0.4)--(2*\hprime + 2*\n-2,-\hprime-0.4); 

\draw[closeInterval,myblue]  (2*\n,-0.4)--(4*\n-1,-0.4);

\draw[shortInterval] (2*\h + 2*\n - 2 + \epsilon,-0.2)--(2*\hprime + 2*\n - 2 - \epsilon,-0.2);

\draw[closeInterval,red] (4*\n - 1,-0.6)--(4*\n - 0.5,-0.6); 

\draw[shortInterval] (4*\n - 1 + \epsilon,-0.5)--(4*\n - 0.5 - \epsilon,-0.5);

      \end{tikzpicture}
    }
    \caption{The closed intervals in ${\cal S}_1\cup \bigcup_{i=1}^4{\cal H}^{2}_i$ of a graph on 4 vertices. We consider $e_1$ to be equal to $v_3v_4$. Each colour represents a different interval size. 
The short intervals are represented by the dots located inside the open (red) intervals.
   Vertical lines mark the endpoints of the intervals in ${\cal S}_1\cup \bigcup_{i=1}^4{\cal H}^{2}_i$,
while the green vertical line marks the beginning of the intervals in ${\cal S}_2$.}\label{fig:intervalCount}
  \end{center}
\end{figure}
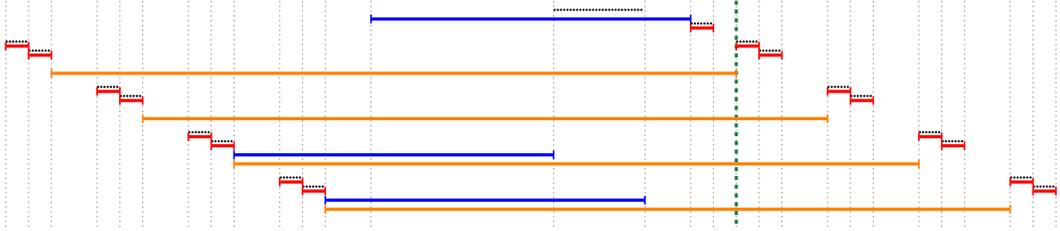

The suitable chosen lengths of the above defined closed intervals are (see~Figure~\ref{fig:intervalCount}, 
where we denote by $\Lambda$
the set of link intervals):
\begin{enumerate}
    \item $0$: short intervals of all grained gadgets (dots in Figure~\ref{fig:intervalCount});
    \item $1/2$: left long and right long  intervals of each ${\cal H}_i^1$, and right long intervals of ${\cal E}_1$ (red intervals in Figure~\ref{fig:intervalCount});
    \item $2n-1$: intervals $C^1_1, \ldots, C^4_1$, and left long intervals of ${\cal E}_1$ (blue intervals in Figure~\ref{fig:intervalCount});
\item $4n-1$: intervals $L^1_i$ and $L^2_i$, for every $i\in [n]$ (orange intervals in Figure~\ref{fig:intervalCount}).
\end{enumerate}

Now, let ${\cal M}'(\mathfrak{G})$ be the interval model where each ${\cal S}_j$ is defined exactly as ${\cal S}_1$,  except that we shift all the intervals to the right in a way that point 0 now coincides with point $t\cdot (j-1)$. More formally, an interval $I$ in ${\cal S}_j$ corresponding to the copy of an interval $[\ell,r]$ in ${\cal S}_1$ is defined as $[\ell+t\cdot (j-1),r+t\cdot(j-1)]$. Also, we assign the intervals in the $(m+1)$-th grained gadgets to be at the end of this model, using the same sizes of intervals as above; i.e., ${\cal H}_i^{m+1}$ is within the interval $[2i-2+t\cdot m,2i-1+t\cdot m]$.

We have shown above that ${\cal M}'(\mathfrak{G})$ has interval count~4.
The following lemma shows that the above chosen intervals satisfy the properties imposed in Subsections~\ref{ss:grai}~and~\ref{ss:redg} on our constructed interval model ${\cal M}(\mathfrak{G})$.

\begin{lemma}
\label{lemma:ic5}
Let $G$ be a cubic graph. Then, there exists an interval model ${\cal M}(\mathfrak{G})$ with interval count~4 for  $\mathfrak{G} = (G,\pi_V,\pi_E)$, for every ordering $\pi_{V}$ and $\pi_{E}$ of the vertex set and edge set of $G$, respectively.
\end{lemma}
\begin{proof}
Denote ${\cal M}(\mathfrak{G})$ by ${\cal M}$. We need to prove that ${\cal M}$ satisfies the conditions of our construction, namely:
\begin{enumerate}
    \item\label{item1} For every $j\in \{1,\ldots,m\}$ and $i\in \{1,\ldots,n\}$, link intervals $L^{2j}_i,L^{2j-1}_i$ weakly intersect ${\cal H}^j_i$ to the right and weakly intersects ${\cal H}^{j+1}_i$ to the left;
    \item\label{item2} For every $j\in \{1,\ldots,m\}$ and $i,i'\in \{1,\ldots,n\}$, $i<i'$, the grained gadget ${\cal H}^{j}_i$ occurs strictly to the left of ${\cal H}^j_{i'}$; 
    \item\label{item3} For every $j\in \{1,\ldots,m\}$, grained gadget ${\cal E}_j$ occurs strictly between the right endpoint of ${\cal H}^j_n$ and the left endpoint of ${\cal H}^{j+1}_1$; and
    \item\label{item4} For every $e_j = v_iv_{i'}\in E(G)$, $i<i'$, intervals $C^1_j,C^2_j$ weakly intersect ${\cal H}^j_i$ to the right and ${\cal E}_j$ to the left, while $C^3_j,C^4_j$ weakly intersect ${\cal H}^j_{i'}$ to the right and strongly intersect ${\cal E}_j$ to the left.
\end{enumerate}

By construction, we know that the right endpoint of ${\cal H}^j_i$ is equal to $2i-1+t(j-1)$, which is also equal to the left endpoints of $L^{2j-1}_i,L^{2j}_i$. Also, the left endpoint of ${\cal H}^{j+1}_i$ is equal to $2i-2+tj$, which is also equal to the right endpoints of $L^{2j-1}_i,L^{2j}_i$ since $t = 4n$; hence Item~\ref{item1} follows. As for Item~\ref{item2}, just note that the right endpoint of ${\cal H}^j_i$, which is equal to $2i-1+t(j-1)$, is strictly smaller than the left endpoint of ${\cal H}^j_{i'}$, which is equal to $2i'-2+t(j-1)$. Indeed, since $i'\ge i+1$, we get $2i'-2\ge 2(i+1)-2 = 2i > 2i-1$. Now, observe that ${\cal E}_j$ is contained in the closed interval $[2n+t(j-1),4n-\frac{1}{2}+t(j-1)]$, that the right endpoint of ${\cal H}^j_n$ is equal to $2n-1+t(j-1)$, and the left endpoint of ${\cal H}^{j+1}_1$ is equal to $tj = 4n+t(j-1)$. Item~\ref{item3} thus follows. Finally, as we have seen, the right endpoint of ${\cal H}^j_i$ is equal to $2i-1+t(j-1)$, which is equal to the left endpoints of $C^1_j,C^2_j$; hence these weakly intersect ${\cal H}^j_i$ to the right. Also, the left endpoint of ${\cal E}_j$ is equal to $2n+t(j-1)$, while the right endpoint of $C^1_j,C^2_j$ is equal to $2(i-1)+2n+t(j-1)$, and all the left short intervals of ${\cal E}_j$ are contained in the open interval $[2(i-1)+2n+t(j-1), 2(i'-1)+2n+t(j-1)]$. Therefore we get that $C^1_j,C^2_j$ weakly intersect ${\cal E}_j$ to the left. Analogously, the right endpoint of ${\cal H}^j_{i'}$ is equal to $2i'-1+t(j-1)$, which is equal to the left endpoints of $C^3_j,C^4_j$; hence they weakly intersect ${\cal H}^j_{i'}$ to the right. Finally, the right endpoint of $C^3_j,C^4_j$ is equal to $2(i'-1)+2n+t(j-1)$, and all the left short intervals of ${\cal E}_j$ are contained in the open interval $[2(i-1)+2n+t(j-1), 2(i'-1)+2n+t(j-1)]$. Also, the left endpoint of the right long intervals of ${\cal E}_j$ is equal to $4n-1+t(j-1)$, which is strictly bigger than $2(i'-1)+2n+t(j-1)$ since $i'\le n$. Therefore, $C^3_j,C^4_j$ strongly intersect ${\cal E}_j$ to the left, finishing the proof of Item~\ref{item4}. 
\end{proof}

We have just shown that, for any orderings $\pi_V$ and $\pi_E$, there exists a model ${\cal M}(\mathfrak{G})$ of 
interval count $4$,
where $\mathfrak{G} = (G,\pi_V,\pi_E)$.
On the other hand, we prove in the remainder of this section that any graph isomorphic to $\mathbb{G}_{{\cal M}(\mathfrak{G})}$ has interval count at least~$4$. 
For this, we show that all such graphs contain as an induced subgraph a certain graph of interval count~$4$, which we denote by $H_4$. 
Next, we define the family $\{H_k\}_{k \geq 2}$ and prove in a more general way that $\mathsf{ic}(H_k) = k$ for every $k \geq 2$. 

\begin{figure}[ht]\centering\captionsetup[subfigure]{justification=centering}
	\begin{subfigure}[b]{0.18\textwidth}\centering
		\includegraphics[scale = 0.68]{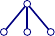}
		\caption{$H_2 = K_{1,3}$}\label{subfig:H_2}
	\end{subfigure}
	\begin{subfigure}[b]{0.26\textwidth}\centering
		\includegraphics[scale = 0.68]{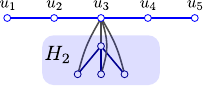}
		\caption{$H_3 = P_{5}\circ H_2$}\label{subfig:H_3}
	\end{subfigure}
\begin{subfigure}[b]{0.26\textwidth}\centering
		\includegraphics[scale = 0.68]{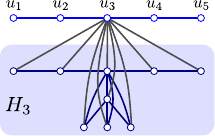}
		\caption{$H_4 = P_{5}\circ H_3$}\label{subfig:H_4}
	\end{subfigure}
	\begin{subfigure}[b]{0.26\textwidth}\centering
		\includegraphics[scale = 0.68]{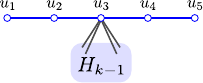}
		\caption{$H_k = P_{5}\circ H_{k-1}$}\label{subfig:H_k}
	\end{subfigure}
	\caption{Graph $H_{k}$ for $k \geq 2$.}
	\label{fig:H_k}
\end{figure}

Let $P_5 = (u_1, \ldots, u_5)$ be a path on $5$ vertices. 
For every graph $H'$, we let $P_5 \circ H'$ be the graph obtained from the disjoint union of $P_5$ with $H'$ by making $u_3$, the central vertex of $P_5$, adjacent to every vertex of $H'$.
In other words, $P_5 \circ H'$ is the graph with vertex set $V(P_5) \cup V(H')$ and edge set $E(P_5) \cup E(H') \cup \{u_3v \mid v \in V(H')\}$.  
Then, for every $k \geq 2$, we let $H_{k}$ be the graph defined recursively as follows (see Figure~\ref{fig:H_k}):
\begin{itemize}
	\item $H_{2} = K_{1,3}$;
	\item $H_{k} = P_{5} \circ H_{k-1}$ for $k > 2$.
\end{itemize}

\begin{lemma}\label{lemma:ic_lower_bound}
	For every $k \geq 2$, $\mathsf{ic}(H_k) = k$. 
\end{lemma}
\begin{proof}
	The proof is by induction on $k$. 
	Since $H_{2} = K_{1,3}$ and $\mathsf{ic}(K_{1,3}) = 2$ c.f.~\cite{roberts1969indifference}, we obtain that the lemma holds for $k = 2$. 
	As inductive hypothesis, suppose that $\mathsf{ic}(H_{k'}) = {k'}$ for some ${k'} \geq 2$. 
	We prove that $\mathsf{ic}(H_{{k'}+1}) = {k'} + 1$.

	First, note that, if ${\cal M}_{P_5} = \{I_1, \ldots, I_5\}$ is an interval model of a $P_5$, 
	with interval $I_i$ representing vertex $u_i$,
	then the precedence relation among the intervals of $I_1, \ldots, I_5$ is either that of Figure~\ref{fig:interval_model_H_k} (i.e., $I_1$ precedes $I_3$, which precedes $I_5$, and $I_2$ precedes $I_4$), or the reverse of the order presented in the figure c.f.~\cite{roberts1969indifference}. 
Let ${\cal M}$ be an interval model of $H_{k'+1}$. 
	Since $H_{k'+1}$ contains a $P_5$ as an induced subgraph, assume without loss of generality that ${\cal M} \supset {\cal M}_{P_{5}}$ and that, with respect to ${\cal M}$, $I_1$ precedes $I_3$, $I_3$ precedes $I_5$, and $I_2$ precedes $I_4$. 
This implies that
	\begin{equation}\label{eq:P5_precedence}
		\ell(I_3) \leq r(I_2) < \ell(I_4) \leq r(I_3)\text{.}
	\end{equation}

	By construction, the only vertex of $P_5$ which is adjacent to the vertices of $H_{k'}$ is its central vertex $u_3$. 
	Consequently, if ${\cal M}_{H_{k'}} \subset {\cal M}$ is the interval model of $H_{k'}$, then there cannot be any intersection between ${\cal M}_{H_{k'}}$ and ${\cal M}_{P_5}\setminus\{I_3\}$, i.e., $I' \cap I_{i} = \emptyset$ for each $I' \in {\cal M}_{H_{k'}}$ and each $i \in \{1,\ldots,5\}$, with $i \neq 3$. 
	Hence, it follows from~\eqref{eq:P5_precedence} that $$\min\{\ell(I') \mid I' \in {\cal M}_{H_{k'}}\} > r(I_2) \text{ and } \max\{r(I') \mid I' \in {\cal M}_{H_{k'}}\} < \ell(I_4)\text{.}$$
	Figure~\ref{fig:interval_model_H_k} illustrates this fact. 
	\begin{figure}[hbt]\centering
		\includegraphics[scale = 0.90]{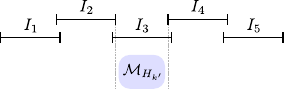}
		\caption{Interval model ${\cal M}_{H_{k'+1}}$ of $H_{k'+1}$.}
		\label{fig:interval_model_H_k}
	\end{figure}
As a result, $I_{3} \supset I'$ for every $I' \in {\cal M}_{H_{k'}}$.
	This, along with the inductive hypothesis that $\mathsf{ic}(H_{k'}) = k'$, implies that $\mathsf{ic}(H_{k'+1}) \geq k'+1$.
	On the other hand, it is straightforward that $\mathsf{ic}(H_{k'+1}) \leq k'+1$ (for instance, consider the model illustrated in Figure~\ref{fig:interval_model_H_k}).
	Therefore, $\mathsf{ic}(H_{k'+1}) = k'+1$. 
\end{proof}

Now, we finally show that $\mathbb{G}_{{\cal M}(\mathfrak{G})}$ contains an $H_4$ as an induced subgraph. 
Since $G$ is cubic, there exists an edge $e_j = (v_{i}, v_{i'}) \in E(G)$ such that $1 < i < i'$.
Let (see Figure~\ref{fig:constructed_graph}):
\begin{itemize}
	\item $I_{1}$ (resp. $I_{2}$) be a right short (resp. long) interval of ${\cal H}_{1}^{j}$;
	\item $I_{3}$ be the link interval $L_{1}^{2j-1}$;
	\item $I_{4}$ (resp. $I_{5}$) be a left long (resp. short) interval of ${\cal H}_{1}^{j+1}$;
	\item $I_{1}'$ (resp. $I_{2}'$) be a right short (resp. long) interval of ${\cal H}_{i}^{j}$;
	\item $I_{3}'$ be the interval $C_{j}^{1}$;
	\item $I_{4}'$ (resp. $I_{5}'$) be a left long (resp. short) interval of ${\cal E}_{j}$;
	\item $J_{1}$, $J_{2}$ and $J_{3}$ be three left short intervals of ${\cal H}_{i+1}^{j}$; and
	\item $J$ be a left long interval of ${\cal H}_{i+1}^{j}$. 
\end{itemize}
The interval graph related to the model comprised by such intervals is isomorphic to $H_4$. 
More specifically, observe first that ${\cal J} = \{J,J_1,J_2,J_3\}$ models $K_{1,3}$. Then, notice that ${\cal P} = \{I_1,\ldots,I_5\}$ and ${\cal P}' = \{I'_1,\ldots,I'_5\}$ model paths on 5 vertices, in this order. Finally observe that $I'_3$ is adjacent to every $I\in {\cal J}$, while there are no edges between ${\cal J}$ and ${\cal P}'\setminus\{I'_3\}$; hence, ${\cal J}\cup {\cal P}'$ is a model for $H_3$. Similarly, $I_3$ is adjacent to every $I\in {\cal J}\cup {\cal P}'$, while there are no edges between ${\cal J}\cup {\cal P}'$ and ${\cal P}\setminus\{I_3\}$; hence ${\cal J}\cup {\cal P}'\cup {\cal P}$ is a model for $H_4$. 
Therefore, $\mathbb{G}_{{\cal M}(\mathfrak{G})}$ has an $H_{4}$ as an induced subgraph, as we wanted to prove.

\section{The interval count of Adhikary et al.'s construction}\label{s:orig}

    We provided in Section~\ref{s:redu} a reduction from the \textsc{MaxCut} problem having as input a cubic graph $G$ into that of \textsc{MaxCut} in an interval graph $G'$ having $\mathsf{ic}(G') \leq 4$. 
Although our reduction requires the choice of orderings $\pi_{V}$ and $\pi_{E}$ of respectively $V(G)$ and $E(G)$ in order to produce the resulting interval model, we have established that we are able to construct an interval model with interval count~$4$ regardless of the particular choices for $\pi_{V}$ and $\pi_{E}$ (Lemma~\ref{lemma:ic5}).
Our reduction was based on that of~\cite{ABMR20}, strengthened in order to control the interval count of the resulting model.

This section is dedicated to discuss the interval count of the original reduction~\cite{ABMR20}. Although the interval count was not of concern in~\cite{ABMR20}, in order to contrast the reduction found there with the one presented in this work, we investigate how interval count varies in the original reduction considering different vertex/edge orderings.
First, we establish that the original reduction yields an interval model corresponding to a graph $G'$ such that  $\mathsf{ic}(G') = O(\sqrt[4]{\lvert V(G')\rvert })$.
Second, we exhibit an example of a cubic graph $G$ for which a choice of $\pi_{V}$ and $\pi_{E}$ yields a model ${\cal M}'$ with interval count $\Omega(\sqrt[4]{\lvert V(G')\rvert})$, proving that this bound is tight for some choices of $\pi_{V}$ and $\pi_{E}$.
For bridgeless cubic graphs, we are able in Lemma~\ref{lemma:ICOrig} to decrease the upper bound by a constant factor, but to the best of our knowledge $O(\sqrt[4]{\lvert V(G')\rvert})$ is the tightest upper bound.
Before we go further analysing the interval count of the original reduction, it is worthy to note that a tight bound on the interval count of a general interval graph $G$ as a function of its number of vertices $n$ is still open. It is known that $\mathsf{ic}(G) \leq \lfloor (n+1)/2 \rfloor$ and that there is a family of graphs $G$ for which $\mathsf{ic}(G) = (n-1)/3$~\cite{COS2012,FIS85}. That is, the interval count of a graph can reach $\Theta(n)$.
    
    In the original reduction, given a cubic graph $G$, an interval graph $G'$ is defined through the construction of one of its models ${\cal M}$, described as follows:
    \begin{enumerate}
        \item let  $\pi_{V} = (v_1,v_2,\ldots,v_n)$ and $\pi_{E} = (e_1,e_2,\ldots,e_m)$ be arbitrary orderings of $V(G)$ and $E(G)$, respectively;
        \item for each $v_i \in V(G)$, $e_j \in E(G)$, let $\mathcal{G}(v_i)$ and $\mathcal{G}(e_j)$ denote respectively a $(p,q)$-grained gadget and a $(p',q')$-grained gadget, where:
            \begin{itemize}
                \item $q = 200n^3 + 1$, $p = 2q + 7n$, and
                \item $q' = 10n^2 + 1$, $p' = 2q' + 7n$;
            \end{itemize}
        \item for each $v_k \in V(G)$, insert  $\mathcal{G}(v_k)$ in ${\cal M}$ such that $\mathcal{G}(v_i)$ is entirely to the left of $\mathcal{G}(v_j)$ if and only if $i < j$. For each $e_k \in E(G)$, insert  $\mathcal{G}(e_k)$ in ${\cal M}$ entirely to the right of $\mathcal{G}(v_n)$ and such that $\mathcal{G}(e_i)$ is entirely to the left of $\mathcal{G}(e_j)$ if and only if $i < j$;
        \item for each $e_j = (v_i,v_{i'}) \in E(G)$, with $i < i'$, four intervals $I^1_{i,j}, I^2_{i,j}, I^1_{i',j}, I^2_{i',j}$ are defined in ${\cal M}$, called \emph{link}  intervals, such that:
        \begin{itemize}
            \item $I^1_{i,j}$ and $I^2_{i,j}$ (resp. $I^1_{i',j}$ and $I^2_{i',j}$) are true twin intervals that weakly intersect $\mathcal{G}(v_i)$ (resp. $\mathcal{G}(v_{i'})$) to the right;
            \item $I^1_{i,j}$ and $I^2_{i,j}$ (resp. $I^1_{i',j}$ and $I^2_{i',j}$) weakly intersect (resp. strongly intersect) $\mathcal{G}(e_j)$ to the left.

\end{itemize}
        By construction, therefore, $I^1_{i,j}$ and $I^2_{i,j}$ (resp. $I^1_{i',j}$ and $I^2_{i',j}$) cover all intervals in grained gadgets associated to a vertex $v_\ell$ with $\ell>i$ (resp. $\ell>i'$) or an edge $e_\ell$ with $\ell < j$.\end{enumerate}
Note that the number of intervals in ${\cal M}$ is independent of what orderings we choose for the vertices and edges of $G$
    and,  therefore, so is the number of vertices of $G'$. 
    Let $n' = \lvert V(G')\rvert $. Since $G$ is cubic, $m = \frac{3n}{2}$. By construction,
    \begin{align*}
      n' & = n(2p+2q)+m(2p'+2q') + 4m = 1200n^4 + 90n^3 + 25n^2 +21n
    \end{align*}
    and thus $n = \Theta(\sqrt[4]{n'})$.
    Since the set of intervals covered by any link interval depends on $\pi_{V}$ and $\pi_{E}$, distinct sequences yield distinct resulting graphs $G'$ having distinct interval counts.
    
    We show next that $\mathsf{ic}(G') = O(\sqrt[4]{n'})$. 
    Note that
    \begin{itemize}
        \item the intervals of all gadgets $\mathcal{G}(v_i)$ and $\mathcal{G}(e_j)$ can use only two interval lengths (one for all short intervals, another for all the long intervals);
        \item for each $e_j = v_iv_{i'} \in E(G)$, with $i < i'$, both intervals $I^1_{i,j}$ and $I^2_{i,j}$ may be coincident in any model, and therefore may have the same length. The same holds for both intervals $I^1_{i',j}$ and $I^2_{i',j}$.
    \end{itemize}
    Therefore, $\mathsf{ic}(G') \leq 2m + 2 = 3n + 2 = \Theta(\sqrt[4]{n'})$. Therefore, the NP-completeness result derived from the original reduction in~\cite{ABMR20} can be strengthened to state that $\textsc{MaxCut}$ is NP-complete for interval graphs $G$ having interval count $O(\sqrt[4]{\lvert V(G)\rvert })$. 
    
\begin{figure}[ht]\centering\captionsetup[subfigure]{justification=centering}
	\begin{subfigure}[t]{0.20\textwidth}\centering
	    \includegraphics[scale=0.65]{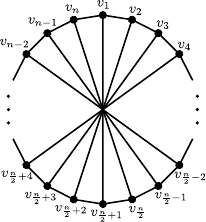}
	    \caption{}\label{fig:arbitraryIC1}
	\end{subfigure}
	\hspace{1.2ex}
	\begin{subfigure}[t]{0.70\textwidth}\centering
	    \includegraphics[scale=0.72]{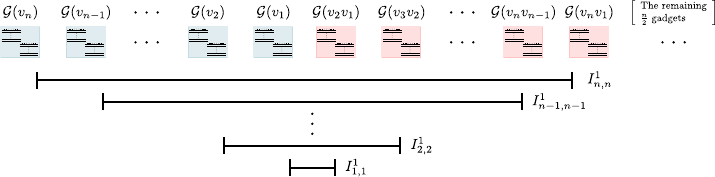}
	    \caption{}\label{fig:arbitraryIC2}
	\end{subfigure}
\caption{(\protect\subref{fig:arbitraryIC1})~A cubic graph $G$, and (\protect\subref{fig:arbitraryIC2})~a chain of nested intervals in the model~${\cal M}'$.}
	\label{fig:arbitraryIC}
	\end{figure}

    Second, we show that there is a resulting model ${\cal M}'$ produced in the reduction, defined in terms of particular orderings $\pi_{V}, \pi_{E}$ for which $\mathsf{ic}({\cal M}') = \Omega(\sqrt[4]{n'})$. Consider the cubic graph $G$ depicted in Figure~\ref{fig:arbitraryIC}(a) which consists of an even cycle $(v_1,v_2,\ldots,v_n)$ with the addition of the edges  $(v_i,v_{i+\frac{n}{2}})$ for all $1 \leq i \leq n/2$. 
For the ordering $\pi_{V} = (v_n,v_{n-1},\ldots,v_1)$ and any ordering  $\pi_{E}$ of the edges starting with the suborder $(v_1v_2, v_2v_3, \ldots, v_nv_1)$ (i.e.  starting with the edges of the cycle), 
the reduction yields a model  ${\cal M}'$ for which there is a chain 
    $I^1_{1,1} \subset I^1_{2,2} \subset \cdots \subset I^1_{n,n}$
    of nested intervals (see Figure~\ref{fig:arbitraryIC}(b)), which shows that $\mathsf{ic}({\cal M}')\geq n$, and thus $\mathsf{ic}({\cal M}') = \Omega(\sqrt[4]{n'})$.

It can be argued from the proof of NP-completeness for \textsc{MaxCut} when restricted to cubic graphs~\cite{BK99} that the constructed cubic graph may be assumed to have no bridges. 
This fact was not used in the original reduction of~\cite{ABMR20}. 
In an attempt to obtain a model ${\cal M}$ having fewer lengths for bridgeless cubic graphs, we have derived Lemma~\ref{lemma:ICOrig}.
Although the number of lengths in this new upper bound has decreased by the constant factor of $4/9$, it is still $\Theta(n) = \Theta(\sqrt[4]{n'})$. 

\begin{lemma}\label{lemma:ICOrig}
Let $G$ be a cubic bridgeless graph with $n = \lvert V(G)\rvert $. There exist particular orderings $\pi_V$ of $V(G)$ and $\pi_E$ of $E(G)$ such that:
\begin{enumerate}
    \item there is a resulting model ${\cal M}$ produced in the original reduction of \textsc{MaxCut} such that $\mathsf{ic}({\cal M}) \leq \frac{4n}{3}+3$. 
    \item for all such resulting models  ${\cal M}$, we have that $\mathsf{ic}({\cal M}) \geq 5$ if $G$ is not a Hamiltonian graph.
\end{enumerate} 
\end{lemma}

\begin{proof}
    Let $G$ be a cubic bridgeless graph with $V(G) = \{v_1,v_2,\ldots,v_n\}$.  
By Petersen's theorem, every cubic bridgeless graph contains a perfect matching, so
    $G$ admits a perfect matching $M$. Let $H = G \setminus M$. Therefore, $H$ is $2$-regular and, therefore, $H$ consists of a disjoint union of cycles $C_1,C_2,\ldots,C_k$, for some $k \geq 1$. For all $1 \leq i \leq k$, let $\pi^i_V = v^i_1, v^i_2, \ldots, v^i_{k_i}$ be an ordering of the vertices of $C_i$, with $k_i = \lvert C_i\rvert$, such that $(v^i_j,v^i_{j+1}) \in E(C_i)$ for all $1 \leq j \leq k_i$, where $v^i_{k_i+1} = v^i_1$. Let $\pi^i_E$ be the ordering $(v^i_1,v^i_2),(v^i_2,v^i_3), \ldots, (v^i_{k_i-1},v^i_{k_i}),(v^i_1,v^i_{k_i})$ for all $1 \leq i \leq k$. 
    Let $\pi_M$ be any ordering of the edges of $M$ such that $(v_i,v_r) < (v_j,v_s)$ in $\pi_M$ only if $v_i < v_j$ in $\pi_V$. 
     Finally, let $\pi_V$ be the ordering of $V(G)$ obtained from the concatenation of the orderings $\pi^1_V, \pi^2_V, \ldots, \pi^k_V$, and  $\pi_E$ be the ordering of $E(G)$ obtained from the concatenation of the orderings $\pi^1_E, \pi^2_E, \ldots, \pi^k_E,\pi_M$. 
    
    In order to prove (2.), assume $G$ is not a Hamiltonian graph. Therefore $k > 1$. Observe that there is the following chain of nested intervals
     $I_1 \subset I_2 \subset I_3 \subset I_4 \subset I_5$, where
    \begin{itemize}
        \item $I_1$ is the leftmost interval in $\mathcal{RS}(\mathcal{G}(v^2_3))$,
        \item $I_2$ is an interval in $\mathcal{RL}(\mathcal{G}(v^2_3))$,
        \item $I_3$ is a link interval corresponding to both $\mathcal{G}(v^2_2)$ and $\mathcal{G}(v^2_1v^2_2)$, 
        \item $I_4$ is a link interval corresponding to both $\mathcal{G}(v^2_1)$ and $\mathcal{G}(v^2_1v^2_{k_2})$, and 
        \item $I_5$ is a link interval corresponding to both $\mathcal{G}(v^1_1)$ and $\mathcal{G}(e)$, where $e$ is the edge of $M$ incident to $v^1_1$,
     \end{itemize}
     since $\ell(I_5) < \ell(I_4) < \ell(I_3) < \ell(I_2) < \ell(I_1) < r(I_1) < r(I_2) < r(I_3) < r(I_4) < r(I_5)$. Thus, for all such resulting models $\mathcal{M}$, we have that $\mathsf{ic}(\mathcal{M}) \geq 5$.
    
    In order to prove (1.), we show that there exists an interval model $\mathcal{M}$, produced by the original reduction of \textsc{MaxCut} considering orderings $\pi_V$ and $\pi_E$, such that $\mathsf{ic}(\mathcal{M}) \leq \frac{4n}{3}+3$, where $n = \lvert V(G)\rvert$. 
    Let $L_1$ be the set of all link intervals of the grained gadgets corresponding to edges of $M$, that is, 
    $L_1 = \{ I^1_{i,k}, I^2_{i,k}, I^1_{j,k}, I^2_{j,k} : e_k = (i, j) \in M \}$. 
    Moreover, let $L_2$ be the set of all link intervals of the grained gadgets corresponding to the edges $(v^i_1, v^i_{k_i})$ of $C_i$ and the vertex $v^i_1$ for all $1 \leq i \leq k$, that is, 
Note that $\lvert L_2\rvert = k \leq n/3$ and $\lvert L_1\rvert = 4 \cdot \lvert M\rvert = 2n$. 
     Let $L = L_1 \cup L_2$. 
     Let $\mathcal{M}' = \mathcal{M} \setminus L$. 
    We claim that $\mathsf{ic}(\mathcal{M}') \leq 3$. 
    Since each pair of true twins $I^1_{j,k}, I^2_{j,k}$ and $I^1_{i,k}, I^2_{i,k}$ in $L_1$ can have the same length in $\mathcal{M}$, it follows from this claim that $\mathsf{ic}({\cal M}) \leq \lvert L_1\rvert+\frac{\lvert L_{2}\rvert}{2} +\mathsf{ic}(\mathcal{M}') \leq \frac{n}{3}+n+3 = \frac{4n}{3}+3$, holding the result. It remains to show that the claim indeed holds.
    
     To prove the claim, let $\mathcal{M}''$ be the interval model obtained from $\mathcal{M}'$ by removing all intervals corresponding to the grained gadgets (or, in other words, by keeping only the intervals corresponding to link intervals). 
     It is easily seen that $\mathcal{M}''$ is a proper interval model, that is, no interval is properly contained in another.
    Therefore, the interval graph corresponding to $\mathcal{M}''$ is a proper interval graph and $\mathcal{M}''$ can be modified so that their intervals have all a single length. 
    Since it is possible to bring all grained gadgets back to $\mathcal{M}''$ using two more lengths, we have that $\mathsf{ic}(\mathcal{M}') \leq 3$, as claimed.
\end{proof}

As a concluding remark,
we note that the interval count of the interval model ${\cal M}$ produced in the original reduction is highly dependent on the assumed orderings of $V(G)$ and $E(G)$, and may achieve $\mathsf{ic}({\cal M}) = \Omega(\sqrt[4]{n'})$. 
The model ${\cal M}'$ produced in our reduction enforces that $\mathsf{ic}({\cal M}') = 4$ which is invariant for any such orderings. On the perspective of the recognition problem for interval graphs with interval count $k$, with fixed $k \ge2$,
for which very little is known, our NP-completeness result on a class of bounded interval count graphs
is also of interest.

\section*{Acknowledgements}
We thank Vinicius F. Santos who shared Reference~\cite{ABMR20}, and anonymous referees for many valuable suggestions, including improving the interval count from $5$ to $4$.

\section*{Data availability statement}
Data sharing not applicable to this article as no datasets were generated or analysed during the current study.

\end{document}